\documentclass[onecolumn,prd,tightenlines,11pt]{revtex4}
\usepackage{amsmath}
\usepackage{amssymb}
\usepackage{graphicx}
\usepackage{epstopdf}
\usepackage{color}

\allowdisplaybreaks[1]

\DeclareGraphicsExtensions{eps,ps}

\newcommand{\comment}[1]{}

\def \beq{\begin{equation}}
\def \eeq{\end{equation}}
\def \beqn{\begin{eqnarray}}
\def \eeqn{\end{eqnarray}}

\def \nn{\nonumber\\}

\def \ket#1{ \left| #1 \right\rangle}
\def \brac#1{\left\langle #1 \right|}

\def \order#1{ {\cal O} \left( #1 \right) }
\def \dslash#1{ #1\!\!\!/}

\def \eps{\varepsilon}
\def \g{\gamma}

\def \CA{{\cal A}}
\def \cP{{\cal P}}

\def \cL{{\cal L}}
\def \cH{{\cal H}}

\def \Op{{\cal O}}

\def \One{\leavevmode\hbox{\small1\kern-3.6pt\normalsize1}} 

\def \MeV{{\rm \; MeV}}
\def \GeV{{\rm \; GeV}}

\def \alS{\alpha_s} 
\def \aS{\alpha_s}       
\def \GF{G_F}              
\def \LamConf{{\Lambda_c}} 

\def \bscc{b\to s + (c\bar{c})}

\def \ccCMF{{c\bar c\,{\rm CMF}}}  
\def \bRF{{b\,{\rm RF}}}           

\def \cc{{c\bar{c}}}               

\def \pp{p_{\perp}}

\def \tr{{\rm Tr}}                 

\def \Ppl#1{{P^+_{#1}}}     
\def \Pmi#1{{P^-_{#1}}}     

\def \xibec{{\xi_{\beta_{\cc},\,\pp}}}        
\def \etaCbec{{\eta^C_{\beta_{\cc},\,\pp}}}  

\def \nbar{{\bar{n}}}

\def \nb {\frac{\dslash{\bar{n}}}{2}}
\def \n {\frac{\dslash{n}}{2}}

\def \Pnb {P_{\bar{n}}}
\def \Pb {P_{\beta_b}}
\def \bp {{\bf p}}

\begin{document}

\title{Decay $b\rightarrow (c\bar{c}) s$ in the leading logarithm approximation.}

\author{Christoph Bobeth,}
\affiliation{Institut f\"{u}r Physik, Universit\"{a}t Dortmund, D-44221~Dortmund, Germany\vspace{4pt} }

\author{Benjamin Grinstein, and Mikhail Savrov}
\affiliation{Department of Physics, University of California at San
  Diego, La Jolla, CA 92093}
  \vspace{5pt}
\date{\today}
\preprint{UCSD/PTH 07-12}

\begin{abstract}
We consider an effective field theory for the nonleptonic decay in which a heavy quark decays into a pair of a heavy quark and antiquark having a small relative velocity and one relativistic (massless) quark. This effective theory is a combination of HQET, SCET, and a covariant modification of NRQCD. In the leading logarithm approximation the effective theory decay amplitude factorizes into the product of matrix elements of heavy-to-heavy and heavy-to-light currents. We discuss a possibility of factorization beyond the leading logarithm approximation and find it doubtful. The Wilson coefficients of the effective theory electro-weak (EWET) Lagrangian in the next-to-the leading logarithm approximation are calculated at the matching scale of the decay. The differential decay rate for the inclusive decay $B\rightarrow J/\psi+h$ in the effective theory framework is evaluated.
\end{abstract}

\maketitle 

\tableofcontents
%

%
\section{Introduction\label{sec:1}}

Decays of $B$ mesons into charmonium are the subject of much theoretical and
experimental study. For example, the CP asymmetry of the decay $B\to J/ \psi K_s$
provides a very high precision determination of the unitarity angle $\beta$.
This is a clean determination because in the Standard Model (SM) of electroweak
interactions the CP asymmetry does not depend on the decay rate. However, it is
an interesting challenge in theoretical physics to determine from first
principles the decay rate in the SM. Doubtless this would shed light into the
underlying QCD dynamics.

We propose to construct an effective field theory that captures the essential
dynamics of decays of a $B$ meson into charmonium plus light mesons. The theory
we propose is a systematic expansion in small parameters and therefore the
resulting approximations can be systematically improved. The building blocks of
the effective theory are largely known: heavy quark effective theory (HQET) to
describe the $b$ quark in the decaying meson~\cite{Isgur:1989vq, Isgur:1989ed,
  Grinstein:1990mj, Georgi:1990um, Falk:1990yz, Falk:1990pz}, soft-collinear
effective theory (SCET) to account for the energetic $s$ quark in the final
state~\cite{Bauer:2000ew, Bauer:2000yr, Bauer:2001ct, Bauer:2001yt, Chay:2002vy,
  Manohar:2002fd}, and non-relativistic QCD (NRQCD) to describe the quark pair
in the charmonium final state~\cite{Caswell:1985ui}-\cite{pa:9910209}. The main adaptation of
these theories to our case involves re-formulating NRQCD covariantly in an
arbitrary frame (normally, NRQCD is formulated in the rest-frame of the
charmonium state). The resulting effective theory, which we call covariant NRQCD
(CNRQCD) accounts for the charm and anti-charm quark fields through four
component spinors, to preserve covariance.

We aim to show that the amplitudes for $B$ decay to charmonium factorize. What
is meant by this is that the amplitude for $B\to \psi + h$, where $\psi$ is a
charmonium state and $h$ the light hadron, is the product of the amplitudes for $B\to h+
J_1$ and $J_2\to \psi$, where $J_{1,2}$ are quark currents. That is, the $B\to
\psi + h$ is given by the form factor of the current $J_1$ between $B$ and $h$
states times the decay constant of the $\psi$ state.  There is a simple physical
picture which suggests factorization. Violations to factorization arise only if
gluons are exchanged between the $\psi$ state and the $B$ or $h$ states. The
effect of very energetic (hard) gluons affects the process at very short
distances only, but will not affect the dynamical picture, so their effect can
be absorbed into an overall constant coefficient. Moreover, this constant is
calculable since hard gluons are perturbative. The effect of low energy gluons,
on the other hand, is non-perturbative. However, the $\psi$ is a very compact
bound state. In a multi-pole expansion, a long wavelength gluon interacts with
the $\psi$ color-charge distribution through it's color-dipole moment since the
state is itself color neutral. But the dipole is very small because the $\psi$
is small. In the theoretical limit of very heavy charm, this coupling to the
dipole vanishes.

It should be noted that the weak interaction can also produce a charm-anti-charm
pair with non-zero total color charge, in the octet configuration. The physical
argument above indicates that in this configuration the quark pair does interact
with soft gluons even at leading order. And while a color octet cannot form a
(color neutral) charmonium state, the emission of a soft gluon can in principle
turn the pair into a color neutral state. It would then seem that this
electroweak contribution to the amplitude for $\psi$ production in $B$ decays
does not factorize. However, a long wavelength gluon interacting with a color
octet leave the state in this octet configuration, and hence in a state which
cannot produce a physical charmonium state. As soon as the gluon wavelength is
short enough to discern the quark-anti-quark nature of the octet, a transition
to the color singlet state can occur. But this is suppressed by the perturbative
coupling constant of short wavelength gluons, so factorization holds at leading order and is expected
to break down at the first order in the small expansion parameters.

While this physical argument is not a proof, it does capture the basic
ingredients needed to construct a more formal argument. As we will see, it will
be important to separate the gluon degrees of freedom according to the
wavelengths and frequencies with which different particles interact. This is
accomplished by using the effective theory approach, combining HQET, SCET and
CNRQCD. Degrees of freedom that are left out in this classification produce what
amount to short distance corrections, which can be absorbed into re-definitions
of coupling coefficients. On the other hand, active degrees of freedom exhibit
some simplifications, as for example spin symmetries, that are of paramount
importance in the solution to the problem.

There is a quite extensive literature on the theory of $B$ decays into
charmonium. The naive factorization hypothesis in exclusive $B\to \psi+h$, with
$\psi$ and $h$ denoting charmonium and strange states, respectively, has been
tested against data, using models of the $B$ to $K$ form factor; see, e.g.,
Refs.~\cite{Gourdin:1994mz, Kamal:1994qb, Colangelo:1995ds, Gourdin:1995zw}.
Several theoretical approaches, distinct from ours, to factorization in QCD have
been applied to exclusive $B$ decays to charmonium. These include ``perturbative
QCD''~\cite{Chou:2001bn} (pQCD) and ``QCD
factorization''~\cite{Beneke:2000ry, Cheng:2001ez}. However, these approaches are
not without trouble.  Ref.~\cite{Song:2002mh} shows that in QCD factorization
infrared divergences arising from nonfactorizable vertex corrections invalidate
factorization.  And the pQCD approach does not find factorization but rather
computes the non-factorizable part.

One may also attempt to formulate factorization for inclusive charmonium
production in $B$ decays, as first proposed by Bjorken~\cite{Bjorken:1988kk} (for semi-inclusive $B \to J/\psi X$ see~\cite{pa:9808360}). This program was formulated more precisely in what is now also commonly known as
NRQCD~\cite{Bodwin:1992qr, Bodwin:1994jh}. This formulation, however, does not
address factorization of amplitudes but rather of decay rates. Moreover, the
program has run into difficulties as predictions of polarization of
charmonium~\cite{Cho:1994ih} have missed the experimental
mark~\cite{Affolder:2000nn}.  Namely, the mechanism which resolves the factor of
30 discrepancy between theoretical predictions and experimental measurements of
prompt $\psi'$ production at the Tevatron involves gluon fragmentation to a
sub-leading Fock component in the $\psi'$ wave-function and yields 100\%
transversely aligned $\psi'$'s to lowest order, which is not observed. The
effective field theory approach presented in this paper may be used to
study inclusive charmonium production in $B$ decays.

The paper is organized as follows. In section~\ref{sec:2} we discuss the kinematics of the decay $B\rightarrow\psi+h$ and outline the effective theory used to derive the decay Lagrangian. In section~\ref{sec:3} we list the degrees of freedom of the effective theory relevant in the leading logarithm  approximation and give the Feynman rules necessary to do one-loop calculations in the effective theory. Also there we discuss the covariant modification of NRQCD. The effective theory electroweak Lagrangian in the leading logarithm approximation and at the leading order in the effective theory power expansion is derived in section~\ref{sec:4}. In section~\ref{sec:5} the simple formal proof of the factorization of the decay amplitude in the leading logarithm approximation is presented. In section~\ref{sec:6} we perform the tree-level matching between the effective theory and full theory amplitudes with the soft (C)NRQCD gluons included and discuss the possibility of factorization beyond the leading logarithms. In Appendix~\ref{sec:11} the differential decay rate for the semi-inclusive decay $B\rightarrow J/\psi+h$ is evaluated. The initial values of the Wilson coefficients in the next-to-leading logarithm approximation at the matching scale of the decay are given in Appendix~\ref{sec:12}. 

%
%
\section{Kinematics \label{sec:2}}

\begin{figure}[htb]
\begin{center}
  \includegraphics[height=3cm]{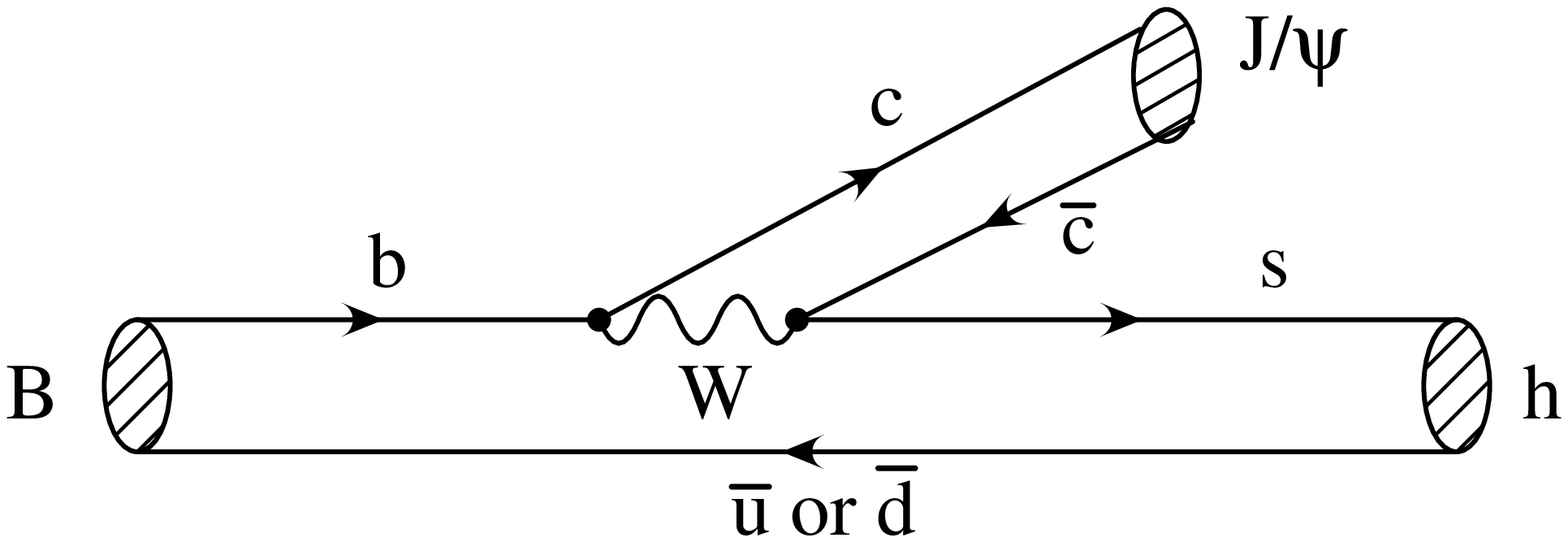}
\end{center}
\caption[The schematic representation of the B-decay]{The schematic
  representation of the decay. The heavy $b$-quark, the heavy quark/antiquark
  $\cc$-pair, the light $s$-quark, and the spectator quark are
  shown.\label{fig:Bdecay}}
\end{figure}

At the quark level the decay $B\to \psi + h$ can be described as the $b$-quark
decaying due to a weak interaction $W$-boson exchange into a $\cc$-pair and the
$s$-quark. The spectator quark in the $B$-meson becomes the second quark in $h$,
see Fig.~(\ref{fig:Bdecay}).

The starting point of our analysis is the effective Lagrangian $\cL_W$ of
$\Delta B = 1$ $B$-decays of electroweak interactions obtained by decoupling
heavy degrees of freedom of the SM such as the $W$-, $Z$-boson and the
$t$-quark at the scale $\mu_W \sim M_W$ of the order of the $W$-boson mass. 
Throughout we will refer to $\cL_W$ as the full theory. The
relevant part which governs at the quark level the decay $\bscc$ is given by
\begin{equation}
  \label{eft:4}
  \cL_W = -\frac{4 \GF}{\sqrt{2}} V^{}_{cb} V^\ast_{cs}
   \Big( C_0( \mu_b ) \Op_0 + C_8( \mu_b ) \Op_8 \Big).
\end{equation}
It has been obtained at the first order in perturbation theory in the
electroweak coupling $\GF$ (Fermi coupling) and the CKM-matrix elements
$V^{}_{cb} V^\ast_{cs}$. The flavor changing $\Delta B = 1$ operators $\Op_i$
are products of two quark currents with the spinor content $[\bar{s} \ldots
b][\bar{c} \ldots c]$ which is the preferable choice when considering the
formation of a $\cc$-bound state as a final state. They read
\begin{align}
  \label{singloct}
  \Op_0 & = [\bar{s} \gamma_\mu P_L b][\bar{c} \gamma^\mu P_L c], &
  \Op_8 & = [\bar{s} \gamma_\mu P_L T^a b][\bar{c} \gamma^\mu P_L T^a c].  
\end{align} 
These operators are related to the electroweak $W$-boson exchange
operators with the ``natural'' spinor content $[\bar{s} \ldots
c][\bar{c} \ldots b]$ by a Fierz transformation.  In the following we
will frequently refer to $\Op_0$ and $\Op_8$ as to the singlet and
octet operators, respectively.

The Wilson coefficients $C_i$ depend on the scale $\mu_b \sim m_b$ which is
of the order of the $b$-quark mass and contain the resummed leading QCD
logarithms of the form $\ln(\mu_b/\mu_W)$ to all orders in $\alS$.  Currently they are known up to the NNLO in QCD \cite{Gorbahn:2004my} in a different operator basis though. The details of
the transformation between two operator bases can be found in
\cite{Gorbahn:2004my} and the exact definition of the evanescent operators
involved in this transformation is given in Appendix~\ref{sec:9}. Here we
give only the numerical values of the coefficients \beq
  \label{eft:7}  
  C_i^{NLO}(\mu) = C_i^{(0)} + \frac{\alS(\mu)}{4 \pi} C_i^{(1)}(\mu).
\eeq

We find for the basis~(\ref{singloct}) at the NLO the central values $C_0^{NLO}
= 0.209$ and $C_8^{NLO} = 2.230$ with the input values~\cite{pa:pdg}: $\alS(M_Z) =
0.1176$, $M_W = 80.403 \GeV$, $M_Z = 91.1876 \GeV$, $m_b(m_b) = 4.2 \GeV\,
(\overline{\rm MS})$, $\mu_W = M_W$ and $\mu_b = m_b$. It should be noted
firstly, that the octet Wilson coefficient is one order larger than that one for the singlet and secondly,
that the NLO correction to the singlet Wilson coefficient is roughly $+50\%$ since at LO:
$C_0^{(0)} = 0.110$ (whereas $C_8^{(0)} = 2.228$).  Furthermore, the scale
uncertainties amount to intervals $C_0^{NLO} \in [0.137, 0.269]$ and $C_8^{NLO}
\in [2.152, 2.343]$ when varying the matching scale of electroweak interactions
$\mu_W \in [55, 115] \GeV$ and the scale $\mu_b \in [2.5, 6.5] \GeV$. This
variation is mainly due to $\mu_b$ dependence which will be cancelled in
physical observables once the matrix elements of the operators are
calculated (up to a higher order $\alS$ residual scale dependence).

Finally, at the scale $\mu_b \sim m_b$ the decay amplitude is proportional to the
matrix element
\begin{equation}
  \label{eft:1}
  \CA( B \to \psi + h ) \sim \brac{\psi,\,h} \cL_W \ket{B}.  
\end{equation}
In the following we will set up the kinematics of the underlying partonic decay
$\bscc$ in the full theory which will enable the construction of an effective theory in
order to further investigate the structure of this matrix element.

The wavelengths of the quarks participating in the decay are reasonably less
than the wavelength associated with the confinement scale $\LamConf\sim 350
\MeV^{-1}$. The $b$-quark and the $\cc$-pair are considered to be heavy since
their masses are roughly one order of magnitude and $3-4$ times larger than the
confinement scale, respectively. The $s$-quark is very light but it is also very
energetic in the restframe of the $b$-quark, so its wavelength is small. The
numerical values of the quark masses are about
\begin{align}
  m_b(m_b) & \sim 4.2  \GeV, &
  m_c(m_c) & \sim 1.25 \GeV, &
  m_s & \sim 0.1  \GeV .
\end{align}
The masses of $b$ and $c$ quarks quoted are obtained from continuum determinations in the $\overline{{\rm MS}}$ scheme~\cite{pa:pdg}. The mass of the $s$-quark is not relevant for our calculation. We will work in the $b$-quark restframe ($\bRF$) and decompose the momenta of the quarks accordingly. The $b$-quark momentum can be decomposed into (omitting
Lorentz indices)
\begin{equation}
  \label{pb:scaling}
  p_b = m_b \beta_b + k_b
\end{equation}
with the scaling of the four velocitiy $\beta_b \sim 1$ and the residual momentum $k_b\sim \LamConf$.

We assume the $\cc$-pair to have a small relative velocity $v = |\vec{v}| \ll c$ in their center of mass frame ($\ccCMF$) in order to form the $\cc$-bound
state $\psi$. In this kinematical configuration they are non-relativistic with a
momentum scaling $\sim m_c (1 + v^2 + \order{v^4}, \vec{v})$. This scaling
persists also to an arbitrary boosted frame
\begin{align}
  \label{pc:scaling}
  p_c & = m_c \beta_\cc + p^\perp_c + k_c, &
  p_{\bar{c}} & = m_c \beta_\cc - p^\perp_c + k_{\bar{c}}
\end{align}
where the four veloctiy of the center of mass of the $\cc$-pair scales as
$\beta_\cc \sim 1$, the perpendicular components as $p^\perp_c \sim m_c v$ and
the residual momenta $k_{c, \bar{c}} \sim m_c v^2$. Throughout the symbol
$v$ will denote the relative velocity of the $\cc$-pair in their $\ccCMF$
whereas four velocities are denoted by $\beta_i$ with $i = \{b, \cc\}$ and $\beta^2_i=1$.

The relative velocity $v$ of the quarks of the $\cc$-pair can be found from the self-consistency condition~\cite{pa:9910209}: 
\beq
\label{scale}
v=\alpha_s(m_c v),\qquad m_c=m_c(m_c)\left[\frac{\alpha_s(m_c v)}{\alpha_s(m_c(m_c))}\right]^{4/\beta_0},
\eeq 
where the first equation follows from the virial theorem applied to the ground state of the hydrogen-like atom: $E_0=-M \alpha_s^2/2=-M v^2/2$ with $M=m_c/2$ is the reduced mass. The second of Eqs.~(\ref{scale}) is the LO expression for the running quark mass in QCD. For 5 flavors $\beta_0=23/3$ and $\Lambda_{QCD}=0.225$ GeV. Numerical solution of Eqs.~(\ref{scale}) gives: $v\approx0.61$, $m\approx1.42$ GeV, so that $m v\approx0.87$ GeV and $m v^2\approx 0.53$ GeV. The relative velocity $v$ is not small although $2m\approx 2.84$ GeV is close to the charmonium mass and $m v^2$ is comparable to $\LamConf$.

Since the effective theory is an expansion in powers of $v$ around $v=0$, the process $\bscc$ at the leading order in $v$-expansion is essentially a two-body decay and we can estimate the energy/momentum of the
outgoing $s$-quark to be $E_s\sim 1.4 \GeV$ in the $\bRF$. Then the energetic
$s$-quark momentum can be decomposed as follows
\begin{equation}
  p_s = (\nbar\cdot p_s)\frac{n}{2} + p_s^\perp + (n \cdot p_s) \frac{\nbar}{2}.
\end{equation}
Here the two vectors $n^\mu = (1, \vec{n})$ and $\nbar^\mu = (1, -\vec{n})$ with $\vec{n}$ being a unit vector (the components are given in the $\bRF$) are light-like vectors, i.e. $n^2 = \nbar^2 = 0$
and $(n\cdot\nbar)=2$. Introducing $p_s^- \equiv (\nbar\cdot p_s)$ and $p_s^+
\equiv (n\cdot p_s)$ the momentum $p_s = (p_s^-, p_s^\perp, p_s^+)$ scales as
$p_s \sim E_s (1, \lambda, \lambda^2)$ with $\lambda \ll 1$ when $\vec{p}_s$ is
chosen to point approximately in the same direction as $\vec{n}$.

So, the wavelengths of all quarks involved are reasonably large compared to $\LamConf$ except for the
spectator quark. Before the decay the spectator quark belongs to the $B$-meson
and after the decay it must become the constituent quark of $h$. 

Temporarily neglecting the spectator quark we will develop an effective theory for the matrix element~(\ref{eft:1}) in the limit where $m_b\rightarrow\infty$,
$m_c\rightarrow\infty$, and $E_s\rightarrow\infty$. We keep the ratio
$r=m_c/m_b$ finite, otherwise the above limits are taken independently. The
latter means that to develop the effective theory for the decay we can
simply combine the leading orders of HQET~(see
  \cite{Grinstein:1990mj}, \cite{Georgi:1990um}, and~\cite{pa:eichill}),
NRQCD~\cite{Bodwin:1994jh}, and SCET~\cite{Bauer:2000yr}.
We want to describe the decay in the $\bRF$, so we have to
develop a covariant formulation of NRQCD.
%
%
\section{Dynamics: HQET, SCET, and CNRQCD\label{sec:3}}

\begin{figure}[hb]
\begin{center}
\includegraphics[height=1.5cm]{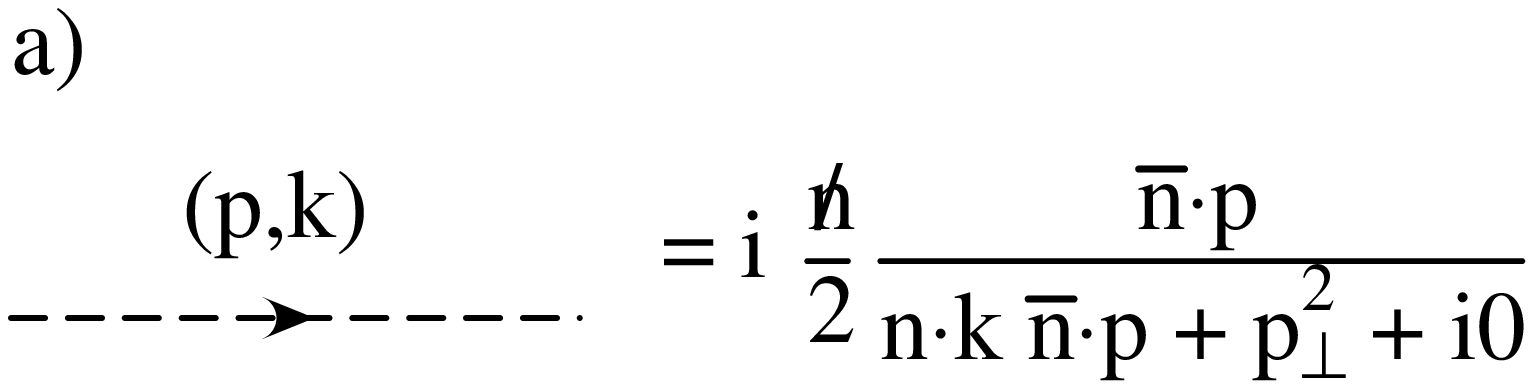}
\includegraphics[height=2cm]{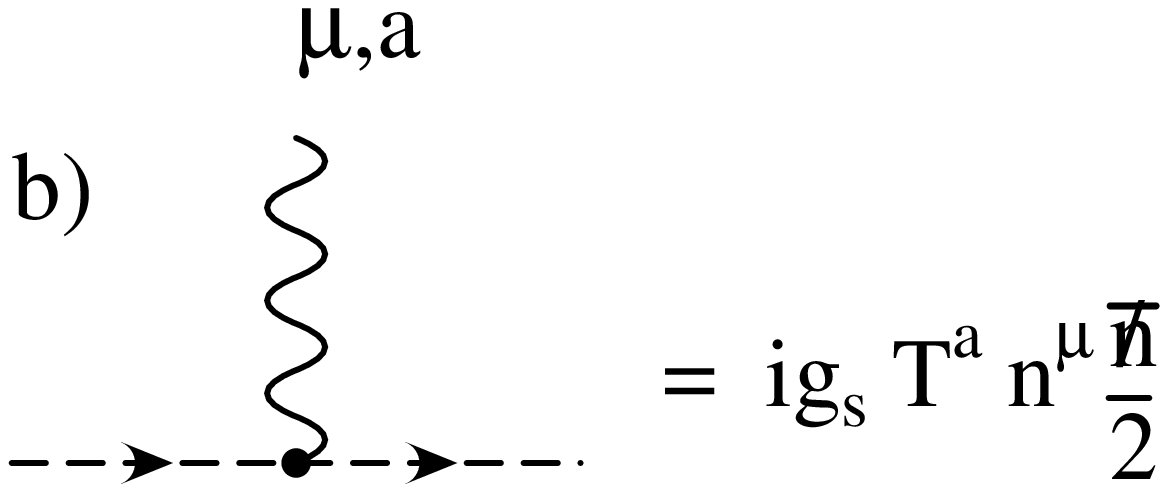}
\end{center}
\begin{center}
 \includegraphics[height=2.2cm]{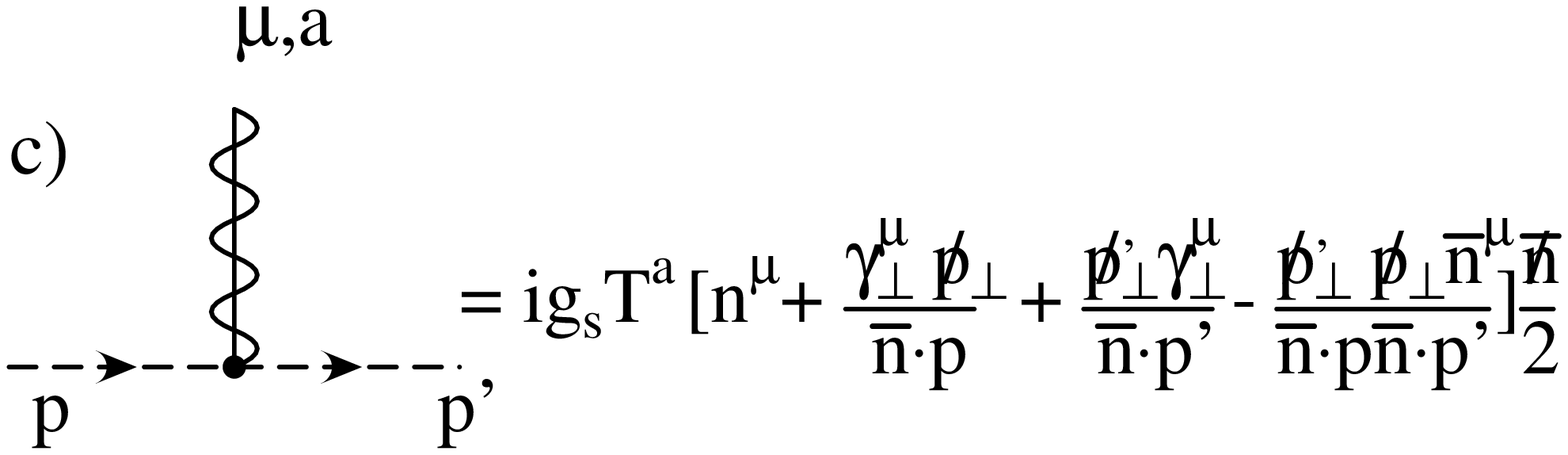}
\end{center}
\begin{center}
  \includegraphics[height=1.5cm]{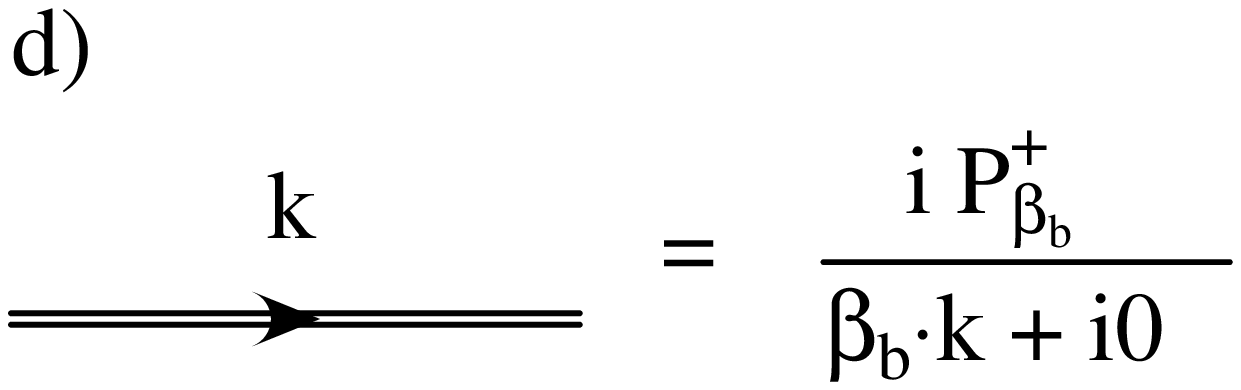}
  \includegraphics[height=2cm]{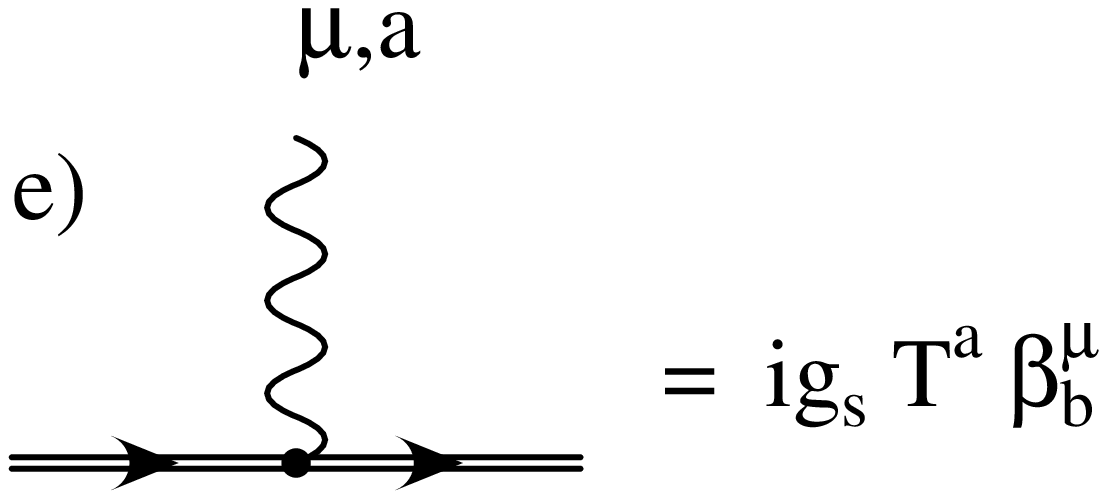}
\end{center}
\begin{center}
 \includegraphics[height=2cm]{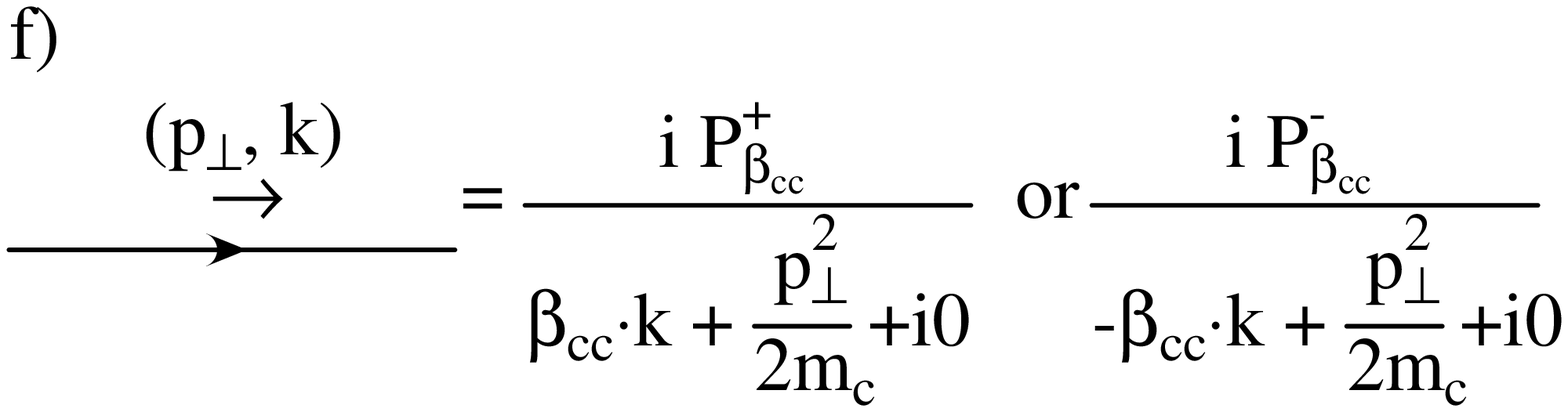}
\end{center}
\begin{center}
 \includegraphics[height=2.3cm]{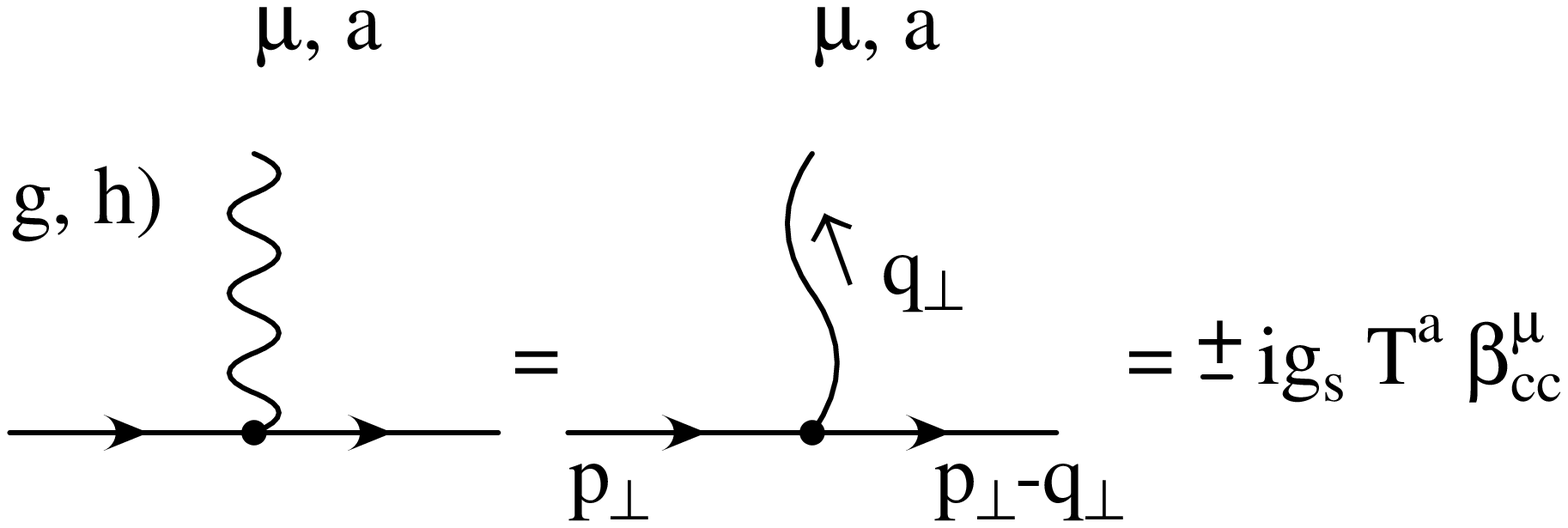}
 \includegraphics[height=1.5cm]{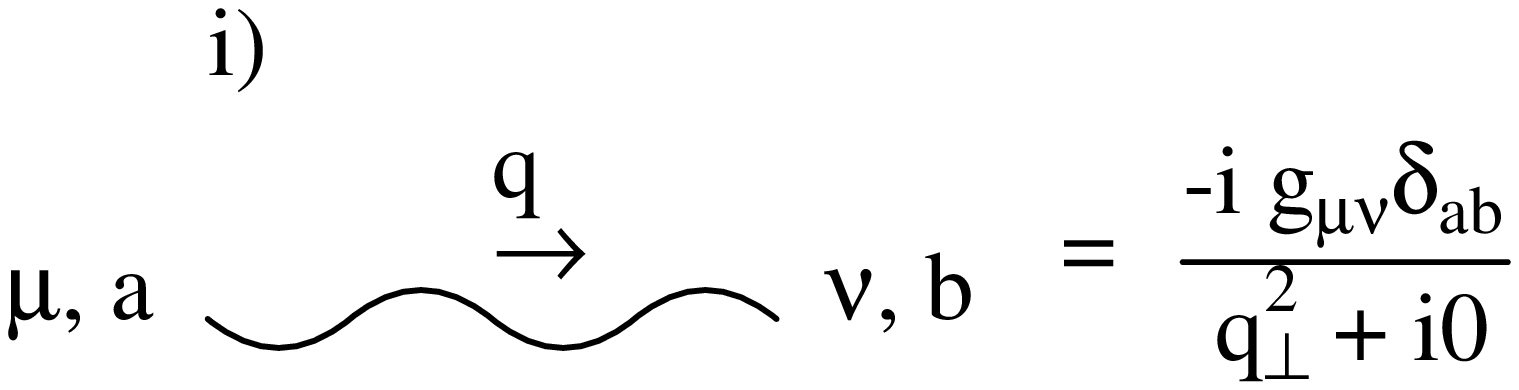}
\end{center}
\begin{center}
  \includegraphics[height=1.5cm]{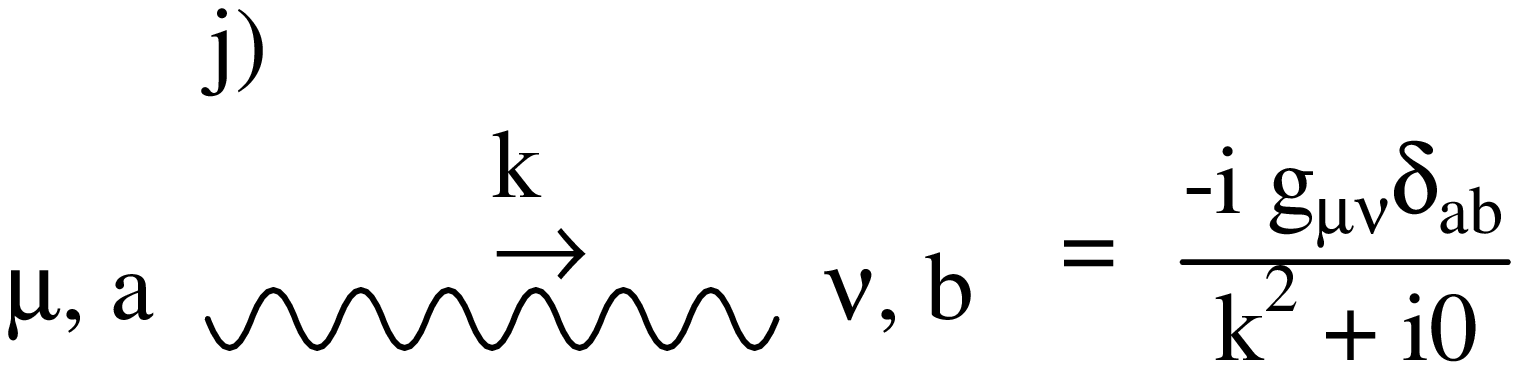}
 \includegraphics[height=1.5cm]{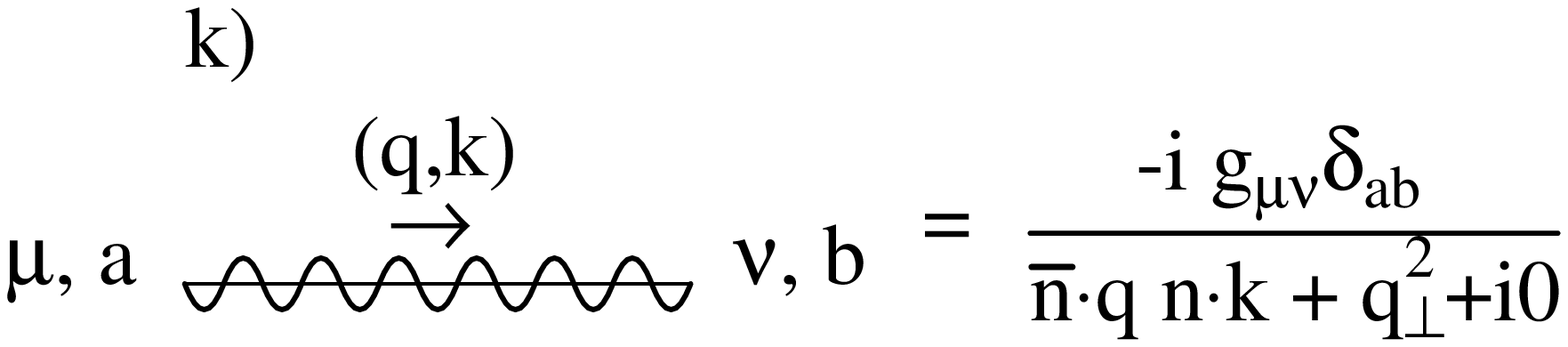}
\end{center}
\caption[Feynman Rules]{\label{FRules} a) Propagator of collinear quark; $n$ is the lightlike vector of the $s$-quark. b) Interaction vertex of the ultrasoft gluon and collinear quark. c) Interaction vertex of the collinear gluon and collinear quark. d) Propagator of the heavy $b$-quark. e) Interaction vertex of the ultrasoft gluon and the heavy $b$-quark. f) Propagator of $c$-quark and/or $\bar{c}$-antiquark in covariant NRQCD; $\beta_{\cc}$ is the 4-velocity of COM of $\cc$ pair. The second expression is the propagator of the charge conjugated field $\eta^C$. g) \& h) Interaction vertex of ultrasoft and/or potential gluon and $c$-quark in covariant NRQCD. Negative sign corresponds to the $\bar{c}$-quark. i) Propagator of the potential gluon in Feynman gauge. Momentum $q_\perp\sim m_c v$ is transverse with respect to the 4-velocity of the $\cc$-pair, $\beta_{\cc}\cdot q_\perp=0$. j) \& k) Propagators of the ultrasoft and collinear gluons in Feynman gauge.}
\end{figure}

In this section we introduce the quark and gluon degrees of freedom of the effective theory and discuss a covariant form of the NRQCD. The Feynman rules necessary for the one-loop calculation in the effective theory are shown in Fig.~\ref{FRules}. We do not discuss HQET and SCET sectors of the effective theory and refer the reader to the extensive literature.
%
\subsection{Field operators\label{fields}}
In short, we need six fields to reproduce the infrared (IR) limit of the full theory amplitude:
\begin{enumerate}
\item 
The outgoing $s$-quark is described by the SCET spinor $\bar{\xi}_{n, \,p}(x)$~(see e.g. \cite{Bauer:2000yr},~\cite{pa:0109045}, and~\cite{pa:0206152}). The large component of the collinear momentum $p=\frac{1}{2}(\bar{n}\cdot p)n+p_{\perp}$ becomes the label $n,\,p$ on the field and the residual $x$-dependence is ultrasoft.
\item
The SCET collinear gluon $A_{n, \,p}^{\mu}(x)$ transferring the collinear momentum. The momentum components scale like those of the collinear quark. 
\item
The incoming $b$-quark is described by the HQET spinor $h_{\beta_b}(x)$~\cite{pa:hqet}.
\item
The outgoing $c$- and $\bar{c}$-quarks are described by CNRQCD spinors $\xi_{\beta_{\cc},\pp}(x)$ and $\eta^C_{\beta_{\cc},\pp}(x)$, respectively. See the next subsection for the discussion.
\item
The ultrasoft field $A_{us}^{\mu}(x)$. This field transfers momentum $\sim m_cv^2\sim E_s\lambda^2\sim\Lambda_c$. 
\end{enumerate}

SCET and NRQCD also include the soft gluons which transfer momenta $\sim E_s\lambda$ and $\sim m_cv$, respectively. At one loop the soft gluons do not contribute to the effective theory decay amplitude, so they're not listed above. The soft gluons are discussed in section~\ref{sec:6} where tree-level matching to all orders in $g_s$ is performed.
%
\subsection{CNRQCD}

\subsubsection{Lagrangian}

In the center-of-mass frame of the $\cc$ pair ($\ccCMF$) the effective Lagrangian is provided by NRQCD. The desired covariant Lagrangian can be obtained from the NRQCD Lagrangian~\cite{pa:griess} in the $\ccCMF$ by a Lorentz boost to the frame where the center-of-mass velocity is $\beta_{\cc}$. At the leading order in $v$-expansion this gives:
\beq
\label{ccLag2}
 \cL_{\cc,\,p} = 
\bar{\xi}_{\beta_{\cc},\pp} 
    \left[ \beta_\cc\cdot iD + \frac{(p^\perp_c )^2}{2 m_c} \right]
    \xi_{\beta_{\cc},\pp} -
  \bar{\eta}^C_{\beta_{\cc},\pp}
  \left[ \beta_\cc \cdot iD 
   - \frac{(p^\perp_c )^2}{2 m_c} \right] \eta^C_{\beta_{\cc},\pp}.
\eeq
Here the covariant derivative $D$ contains only the ultrasoft field $A_{us}$. When calculating one-loop QCD corrections to the effective theory Lagrangian it is more convenient to work with the charge conjugated field $\etaCbec$ because it creates the anti-quark $\bar{c}$ in the final state and this is exactly what we
need in the effective theory for the $B\to \psi + h$ decay.

At the leading order in $v$-expansion the effective theory Dirac spinors $\xi_{\beta_\cc,\,{\pp}}$ and $\etaCbec$ are related to the full theory field operator $\psi$ as:
\beq
\label{operators}
\psi(x)=e^{-im_c\beta_{\cc}\cdot x}\sum_{{\pp}\neq0}e^{-i {\pp}\cdot x}\xi_{\beta_\cc,\,{\pp}}(x)+e^{im_c\beta_{\cc}\cdot x}\sum_{\pp\neq0}e^{i \pp\cdot x}\eta^{C}_{\beta_\cc,\pp}(x)+\order{v},
\eeq
with $\pp$ the covariant potential labels. We use symbols $\xi$ and $\eta$ for the quark and anti-quark fields following the conventional NRQCD notation. However unlike in the standard approach we keep all four components of the spinors to ensure relativistic covariance. The 4-spinors satisfy the constraints
\begin{align}
  \label{const}
  \Ppl{\beta_\cc} \xibec &= 0, &
  \Ppl{\beta_\cc} \eta_{\beta_{\cc},\,\pp} & = 0.
\end{align}
Here $\Ppl{\beta}=(1+\dslash{\beta})/2$. The solutions of (\ref{const}) are such that in the $\ccCMF$ their upper components become the conventional NRQCD two-component Pauli spinors $\xi_{\bf p}$ and $\eta_{\bf p}$ and the lower components vanish:
\begin{align}
  \label{xixi}
  \xibec & = \sqrt{\frac{1 + \beta_\cc^0}{2}} \left(
    \begin{array}{c} \xi_{\bf p} \\ 
      \frac{\vec{\beta}_\cc \vec{\sigma}}{1+\beta^0_\cc}\xi_{\bf p} \end{array}
  \right), &
  \eta_{\beta_{\cc},\,\pp} & =\sqrt{\frac{1+\beta_\cc^0}{2}} \left(
    \begin{array}{c} \eta_{\bf p} \\ 
      \frac{\vec{\beta}_\cc \vec{\sigma}}{1+\beta^0_\cc}\eta_{\bf p} \end{array}
  \right).
\end{align}

In other words $\xibec$ and $\eta_{\beta_{\cc},\,\pp}$ are the NRQCD spinors
$\xi_{\bf p}$ and $\eta_{\bf p}$ boosted into the frame with velocity
$-\vec{\beta}_\cc (\beta^0_\cc)^{-1}$ with respect to the $\ccCMF$. In the
$\ccCMF$ $\beta_\cc = (1,{\bf 0})$ and the standard form of the NRQCD
Lagrangian is reproduced. This formalism is essentially the same as the one discussed in
\cite{pa:spinors}. The spinor  $\etaCbec$ satisfies 
\beq\label{eq14}
\Pmi{\beta_\cc} \etaCbec = \etaCbec,
\eeq
where $\Pmi{\beta}=(1-\dslash{\beta})/2$, and transforms under the same representation of $SU(3)$ as $\xibec$. The explicit form of $\etaCbec$ in terms of Pauli spinor $\eta_{\bf p}$ is given by~(\ref{vspin3}).

NRQCD also includes the off-shell quark modes called soft quarks. The energy and momentum of the soft quark both scale like $\sim m_cv$ and therefore the Lagrangian for the soft quark at the leading order in $v$-expansion is the same as HQET Lagrangian. However we will see in section~\ref{sec:4} that at one loop the soft gluons do not contribute to the decay amplitude and so we do not discuss them here.

The NRQCD part of the decay amplitude also includes a $1/v$ (Coulomb) singularity. This singularity is reproduced by the local (on the ultrasoft scale) four-quark operator~(see~\cite{pa:9912226} and~\cite{pa:9910209}) corresponding to the Coulomb interaction between the heavy quarks. In this paper instead we have adopted a simplistic approach: we introduce the potential gluon that transfers the frequency $\sim m_c v^2$ and momentum $\sim m_cv$ (in the $\ccCMF$) and does not remove the quarks off-shell~\cite{pa:griess}. This is enough to reproduce the Coulomb singularity at one loop. The Lagrangian that couples the potential gluon and the on-shell quarks changes the momentum label $\pp$ on the quark fields $\xi$ and $\eta^C$ and gives rise to the same interaction vertex as the ultrasoft gluon.

\subsubsection{Full theory spinors in terms of $\xi$ and $\eta$}

Matching a full QCD matrix element onto the matrix element of the CNRQCD requires writing down
the full QCD spinors $u$ and $v$ in terms of the CNRQCD spinors $\xi$ and $\eta$. This is done as in NRQCD only the frame of reference is not specified. For example:
\beqn
  \label{uspin}
  && (m\dslash{\beta}+\dslash{p}_c-m)u = 0, \nn
  && (-2mP^-_{\beta}+\dslash{p}_c)u = 0, \nn
  && (-2mP^-_{\beta}+P^-_{\beta}\dslash{p}_c)(P^+_{\beta}+P^-_{\beta})u = 0, \nn
  && (-2mP^-_{\beta}+P^-_{\beta}\dslash{p}_{c\perp}P^+_{\beta}-\beta\cdot p_c P^-_{\beta})u = 0, \nn
  && P^+_{\beta}u = \xi_{\beta_{\cc},\,\pp},\qquad P^-_{\beta}u = u_-, \qquad u = \xi_{\beta_{\cc},\,\pp}+u_-,\nn
  && -(2m+\beta\cdot p_c)u_-+P^-_{\beta}\dslash{p}_{c\perp}\xi_{\beta_{\cc},\,\pp} = 0, \nn
  && u_-=\frac{1}{2m+\beta\cdot p_c}P^-_{\beta}\dslash{p}_{c\perp}\xi_{\beta_{\cc},\,\pp},\nn
  && u = \left[ 1 + \frac{1}{2m+\beta\cdot p_c}P^-_{\beta}\dslash{p}_{c\perp}\right] 
      \xi_{\beta_{\cc},\,\pp} \stackrel{\beta=(1,{\vec 0})}{\rightarrow} 
     \left( \begin{array}{c} \xi_{\bf p} \\
            \frac{\vec{p}\vec{\sigma}}{E+m}\xi_{\bf p} \end{array} \right).
\eeqn
Here $p_c$ denotes the sum of the perpendicular and residual components in the momentum decomposition given in~(\ref{pc:scaling}). In the equation the effective theory spinor $\xi_{\beta_{\cc},\,\pp}$ is introduced as a projection of the full QCD spinor $u$ onto the subspace restricted by the
constraint~(\ref{const}). The second half of the bispinor subspace $u_-$
accounts for the difference of the order $v$ between the full and effective theories. One can see
that in the $\ccCMF$ the full QCD spinor is reproduced. Unlike the $\xi$-spinor
which has only two upper components in the $\ccCMF$, the full QCD spinor has
also two non-zero lower components suppressed by the first power of velocity in
the $\ccCMF$.

To parameterize the $v$-spinor of the full theory by the $\eta^C$-spinor notice that the charge conjugated spinor $v^c=Cv^\ast$ satisfies the same equation as the $u$-spinor. The CNRQCD
Lagrangian for the anti-quark is derived from the full theory using charge conjugation. Therefore the 
$v^c$-spinor must be written in terms of the effective theory $\eta$-spinor in the same way as the
$u$-spinor is written in terms of the $\xi$-spinor and then the $v$-spinor follows after charge
conjugation:
\beqn
  \label{vcspin}
  v= i\gamma^2{v^{c}}^{\ast} && 
    = i\gamma^2 \left[1 + \frac{1}{2m+\beta\cdot p_c}P^-_{\beta}\dslash{p}_{c\perp} \right]^\ast
      \eta^\ast_{\beta_{\cc},\,\pp}, \nn
  && = \left[1 - \frac{1}{2m+\beta\cdot p_c} P^+_{\beta}\dslash{p}_{c\perp}\right] 
        (i\gamma^2\eta^\ast_{\beta_{\cc},\,\pp}), \nn
  && = \left[1 - \frac{1}{2m+\beta\cdot p_c} P^+_{\beta}\dslash{p}_{c\perp}\right]\eta^C_{\beta_{\cc},\,\pp}.
\eeqn
Here, componentwise:
\beq
  \label{vspin3}
  \eta^C_{\beta_{\cc},\,\pp} = (i\gamma^2\eta^\ast_{\beta_{\cc},\,\pp})
   = \left( \begin{array}{c} \frac{\vec{\beta}\vec{\sigma}}{1+\beta^0}(-i\sigma^2\eta_{\bf p}^\ast) \\ 
            (-i\sigma^2\eta_{\bf p}^\ast) \end{array} \right)
  \qquad{\rm where}\qquad 
  (-i\sigma^2\eta^\ast) = (-\eta^\dagger_{\downarrow},\eta^\dagger_{\uparrow})^T.
\eeq
Notice that the spinor $\eta^C_{\beta_{\cc},\,\pp}$ satisfies~(\ref{eq14}), i.e. for
the charge conjugated spinor the 4-velocity is reversed.

%
%
\section{EWET Lagrangian in the leading logarithm approximation\label{sec:4}}

To obtain the effective theory electroweak (EWET) Lagrangian we perform the tree-level matching between the full theory amplitude of $b\rightarrow(\cc)s$ decay and the decay amplitude in the effective theory to all orders in $g_s$ and at the leading order in $v$ and $\lambda$. Then the result is RG-improved using the anomalous dimension matrix extracted from the results of one-loop matching which corresponds to summing the leading Sudakov logarithms and gives the EWET Lagrangian in the leading logarithm approximation. The soft gluons do not contribute at one loop and therefore do not modify the anomalous dimension matrix in the leading logarithm approximation. From this we infer that in this approximation the soft gluons can be ignored.
%
\subsection{Tree-level matching}

\begin{figure}[h]
\begin{center}
  \includegraphics[height=2.7cm]{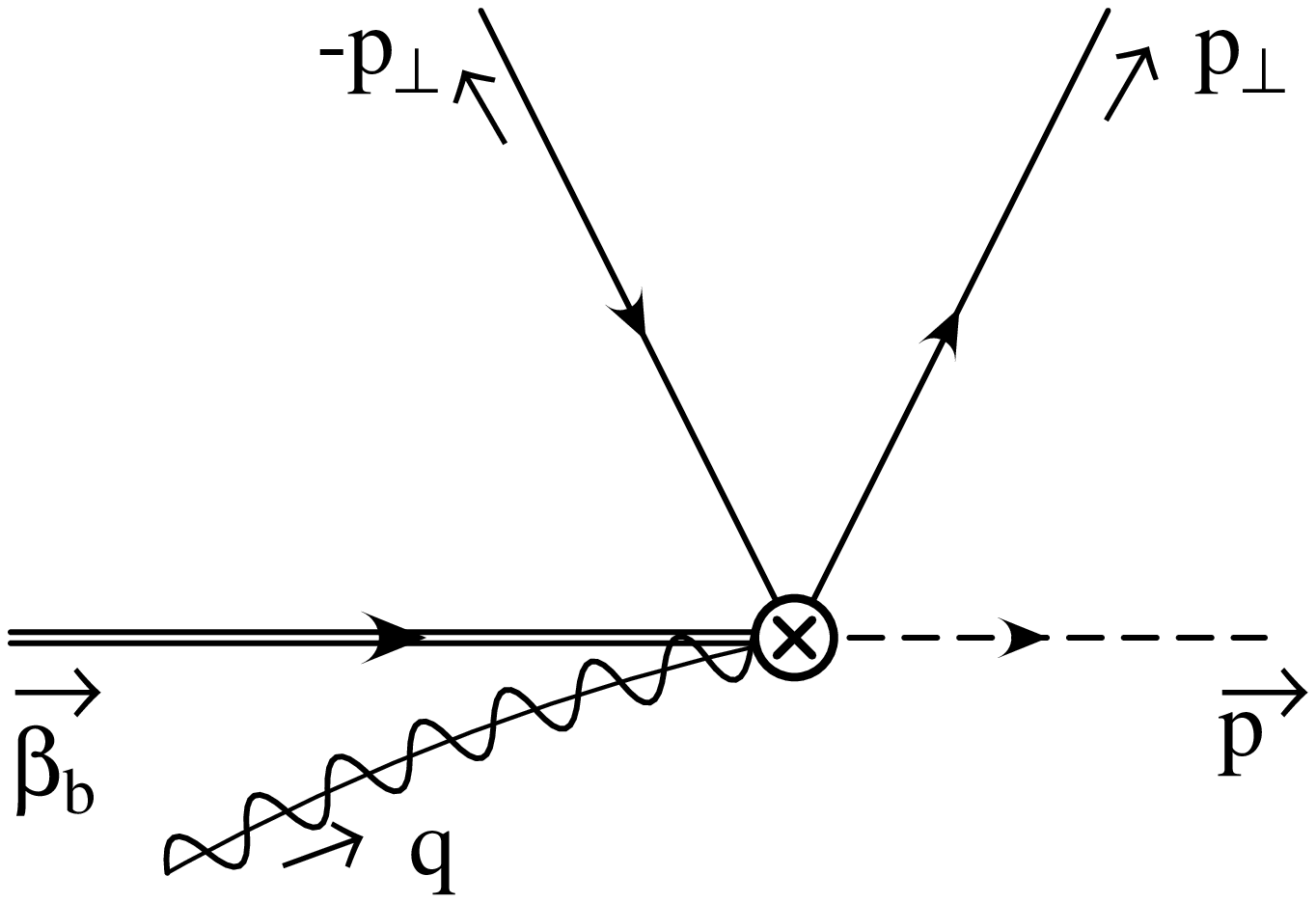}
  \caption[ET amplitude at the tree-level]{The effective theory amplitude at the tree-level. Double line corresponds to the incoming heavy $b$-quark (labeled by 4-velocity $\beta_b$), the dashed line corresponds to the outgoing light $s$-quark (labeled by collinear momentum $p$), the solid lines correspond to the outgoing quarks of $\cc$-pair (labeled by the potential momenta $\pm\pp$). For the $\bar{c}$ anti-quark the fermion flow is reversed as in the full theory. Only one collinear gluon incoming to the electro-weak vertex is shown.\label{fig:4}}
  \end{center}
\end{figure}

At tree level the matching is done by replacing the full theory fields of~(\ref{eft:4}) with the effective theory fields of section~\ref{fields} and using the reduction formula (see Appendix~\ref{sec:10})
\beq
\label{ETstruct1}
 \g_\mu P_L\otimes \g^\mu P_L\rightarrow \frac{1}{2}P_R\otimes \dslash{n}_\perp
  -\frac{1}{2}(\beta_{\cc}\cdot n)P_R\otimes \g_5 - \frac{1}{2}\dslash{\eps}_-\otimes \dslash{\eps}_{+\perp}
\eeq
to decompose the full theory Dirac structure into the sum of the three Dirac structures of the effective theory. As we have inferred the collinear gluons are the only degrees of freedom giving rise to the local operators on the effective theory side in the leading logarithm approximation. Then the result of tree-level matching follows easily from the $\order{g_s}$ calculation and the SCET invariance of the effective theory Lagrangian that requires the collinear gluons and the collinear quark fields to combine in the jet field~\cite{Bauer:2001ct}: 
\beq
\label{EWETmatch2}
\Op^{(\text{tree})}=
    [\bar{\xi}_{n,p}\,W\,\Gamma_j {\bf C}\,h_{\beta_b}]
   [\bar{\xi}_{\beta_{\cc}\,\pp}\,\Gamma_j{\bf C}\,\eta^C_{\beta_{\cc}\,-\pp}].
\eeq
Here ${\bf C}=\One\,\,{\rm or}\,\,T^a$ and $\Gamma_j \otimes \Gamma_j = \{ 
P_R \otimes \dslash{n}_\perp, \,
P_R \otimes \g_5,\,
\dslash{\eps}_{-} \otimes \dslash{\eps}_{+\perp}
 \}$. The sum over potential label $\pp$ is understood. The operator $W$ is the Wilson line
\beq
\label{wl}
W=\sum_{\text{perm}}\exp\left(-g_s\frac{1}{\bar{n}\cdot \cP}\bar{n}\cdot A_{n,\,q} \right)
\eeq
where $\cP$ is the operator that picks up the net label momentum on the right. Momentum conservation implies that at each order in $g_s$ the labels on the collinear fields satisfy $p+\sum_iq_i=m_b\beta_b-2m_c\beta_{\cc}$. The corresponding Feynman diagram in the effective theory is shown in Fig.~\ref{fig:4}. 
%
\subsection{Decay amplitude at $\order{\aS}$}

Here we present the results of one-loop calculation of the decay amplitude in the effective and full theories. Feynman diagrams are calculated in the $\overline{\text{MS}}$-scheme. The momenta of the four external particles are set on-shell and the corresponding IR divergences are regulated by the dimensional regularization. 
%
\subsubsection{Effective theory\label{ETamplitude}}

Overall in the effective theory there are eight amputated diagrams. Seven of them are due to loop corrections to the zero order term in the $g_s$ expansion of the tree-level Lagrangian~(\ref{EWETmatch2}):
\beqn
&&\Op^{(0)}_{i\,j}= [\bar{\xi}_{n,p}\Gamma_j {\bf C}_i\, h_{\beta_b}]
   [\bar{\xi}_{\beta_{\cc}\,\pp}\Gamma_j{\bf C}_i\,\eta^C_{\beta_{\cc}\,-\pp}],
   \label{gamma:1}
\eeqn
where $i = 0, 8$ stands for the singlet and octet operators and
   $j = 1,2,3$ for the Dirac structures introduced in~(\ref{ETstruct1}).

\begin{figure}[htb]
\begin{center}
\begin{minipage}[h]{5cm}
  \includegraphics[height=1.5cm]{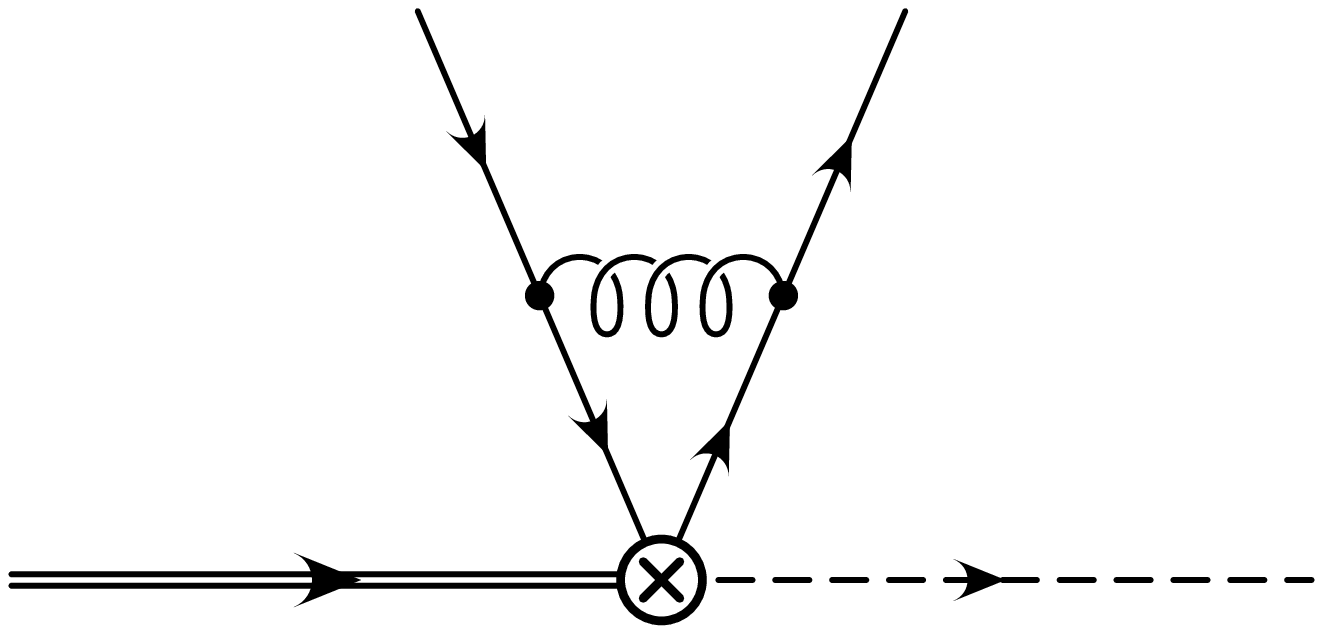}
\end{minipage} 
\begin{minipage}[h]{5cm}
  \includegraphics[height=1.5cm]{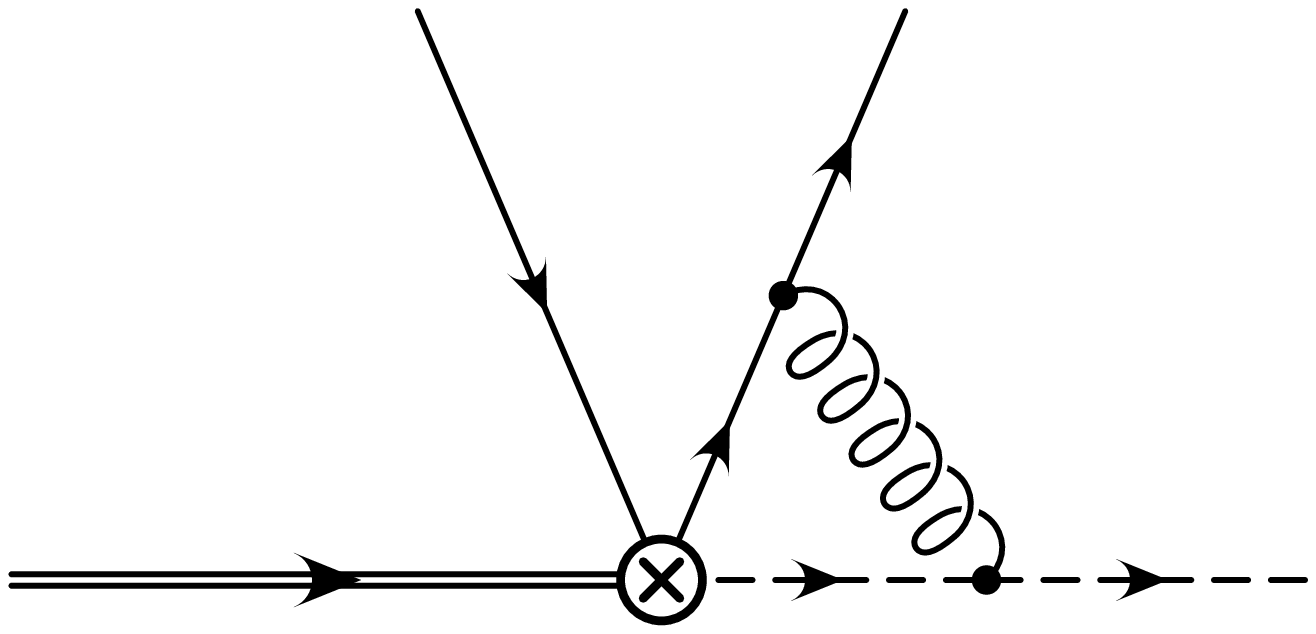}
\end{minipage}
\begin{minipage}[h]{5cm}
  \includegraphics[height=1.5cm]{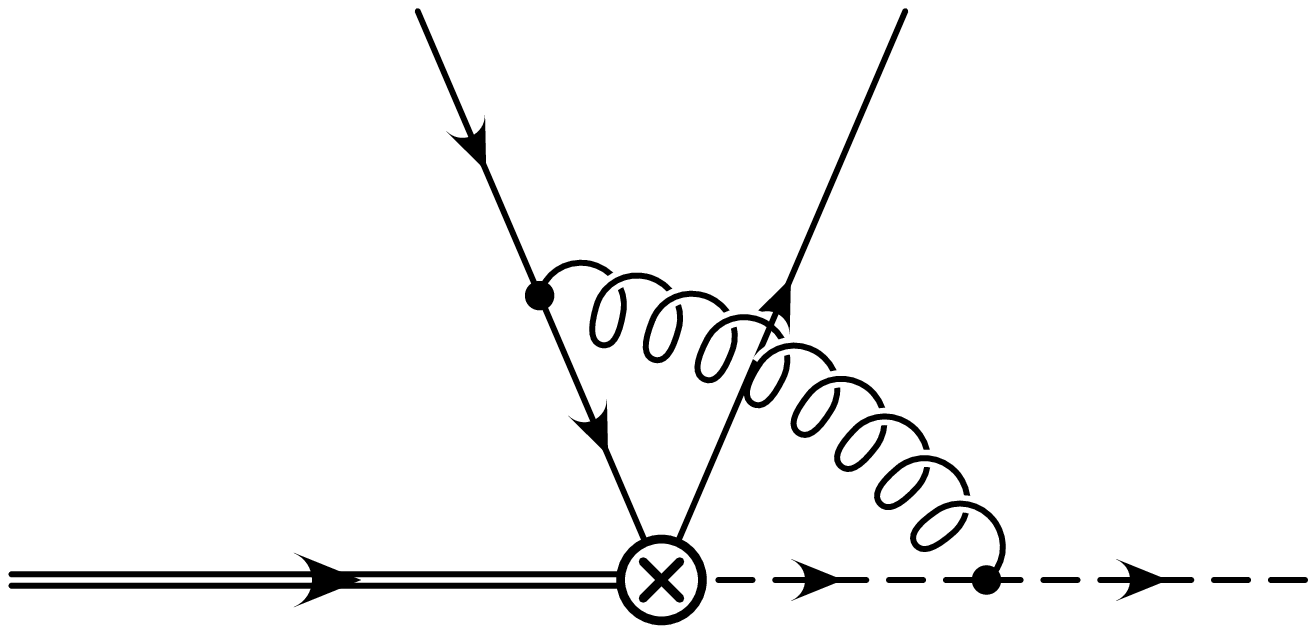}
\end{minipage}
\begin{minipage}[h]{5cm}
  \includegraphics[height=1.5cm]{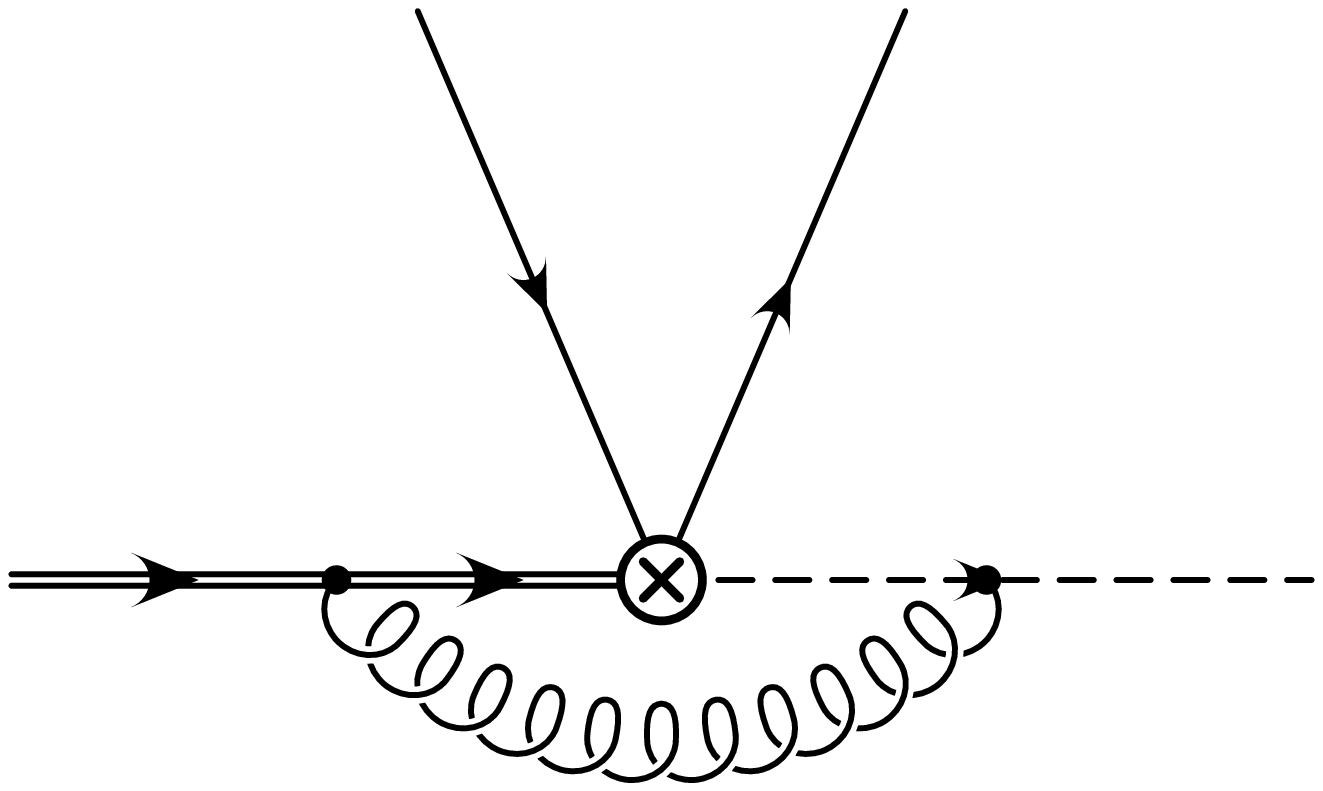}
\end{minipage}
\begin{minipage}[h]{5cm}
  \includegraphics[height=1.5cm]{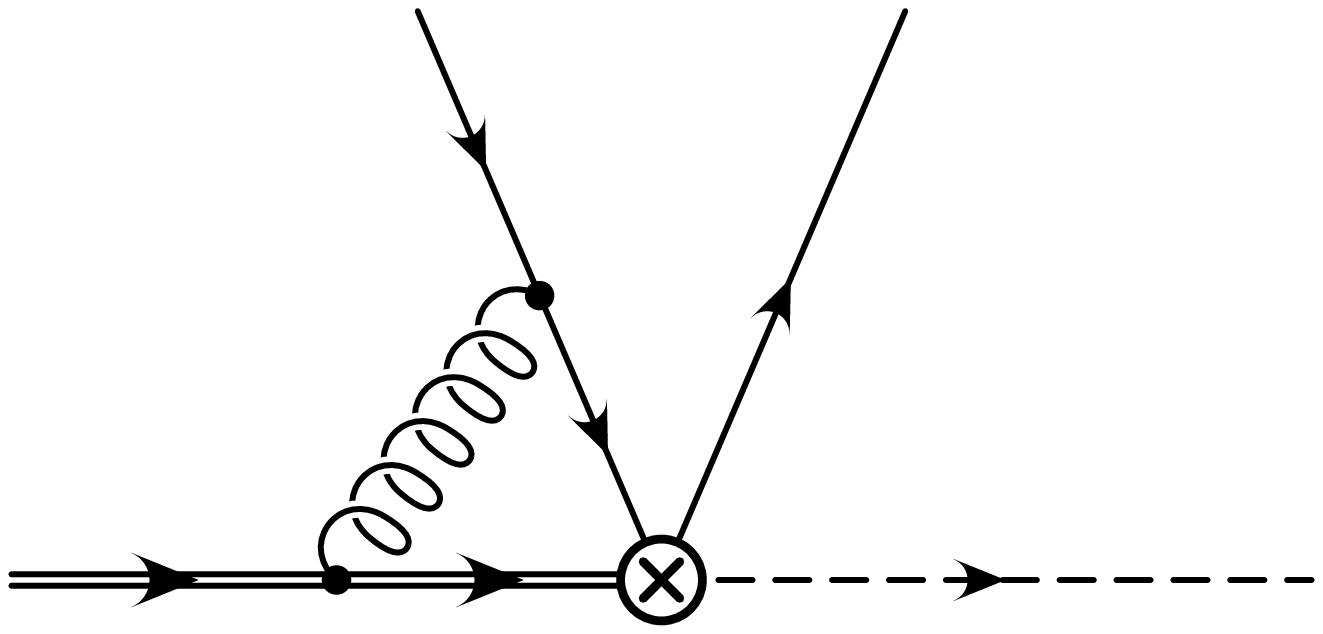}
\end{minipage} 
\begin{minipage}[h]{5cm}
  \includegraphics[height=1.5cm]{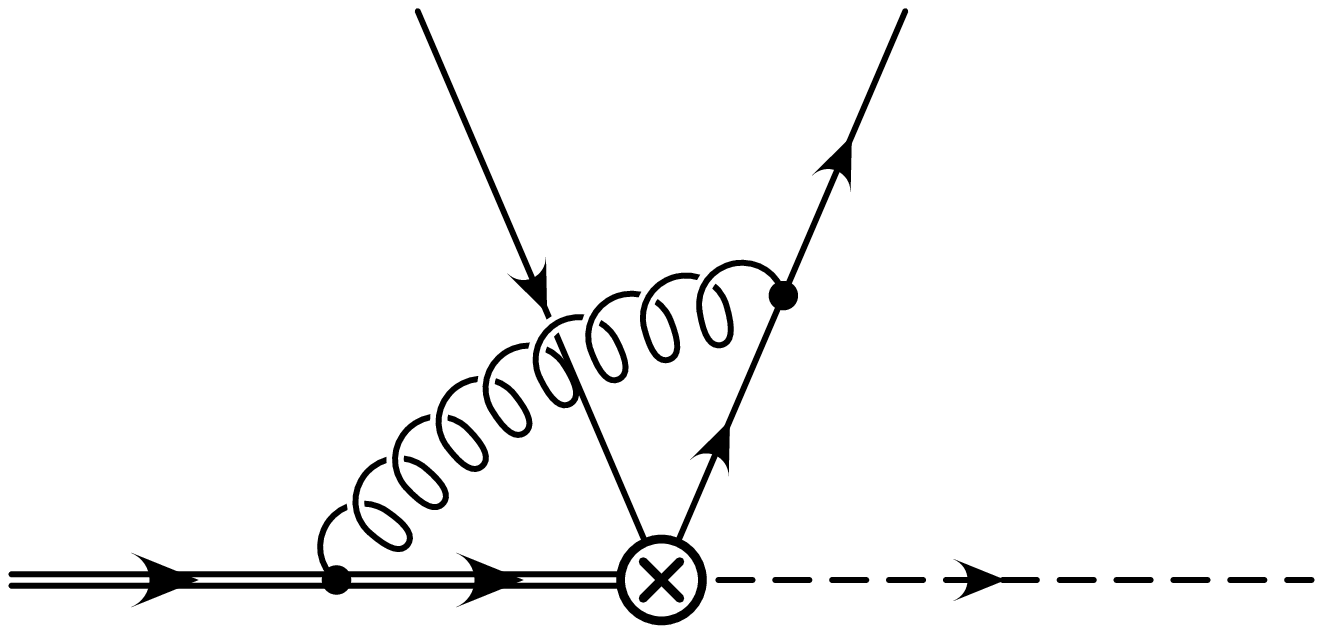}
\end{minipage}
\end{center}
  \caption[Diagrams with the ultrasoft gluon exchange in ET]{\label{fig:usoft}Diagrams with the ultrasoft gluon exchange in the effective theory.}
\end{figure}

\begin{figure}[htbp]
\begin{center}
\begin{minipage}[h]{5cm}
 \includegraphics[height=1.5cm]{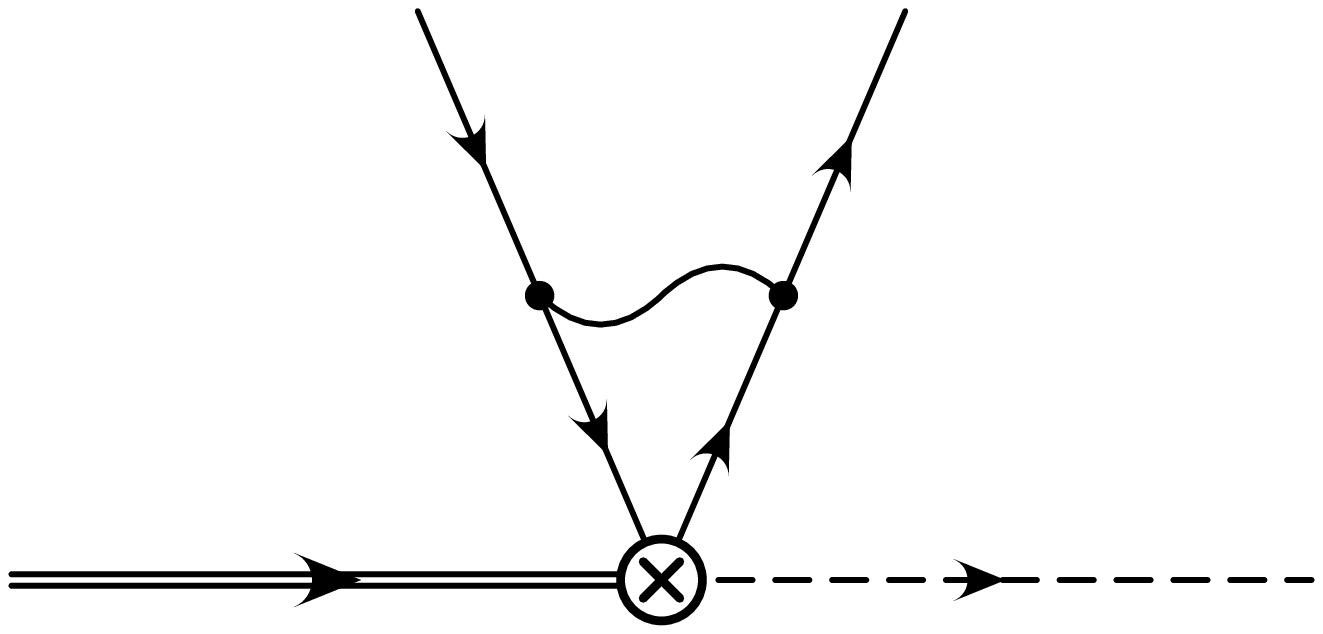}
 \end{minipage}
 \begin{minipage}[h]{5cm}
  \includegraphics[height=2cm]{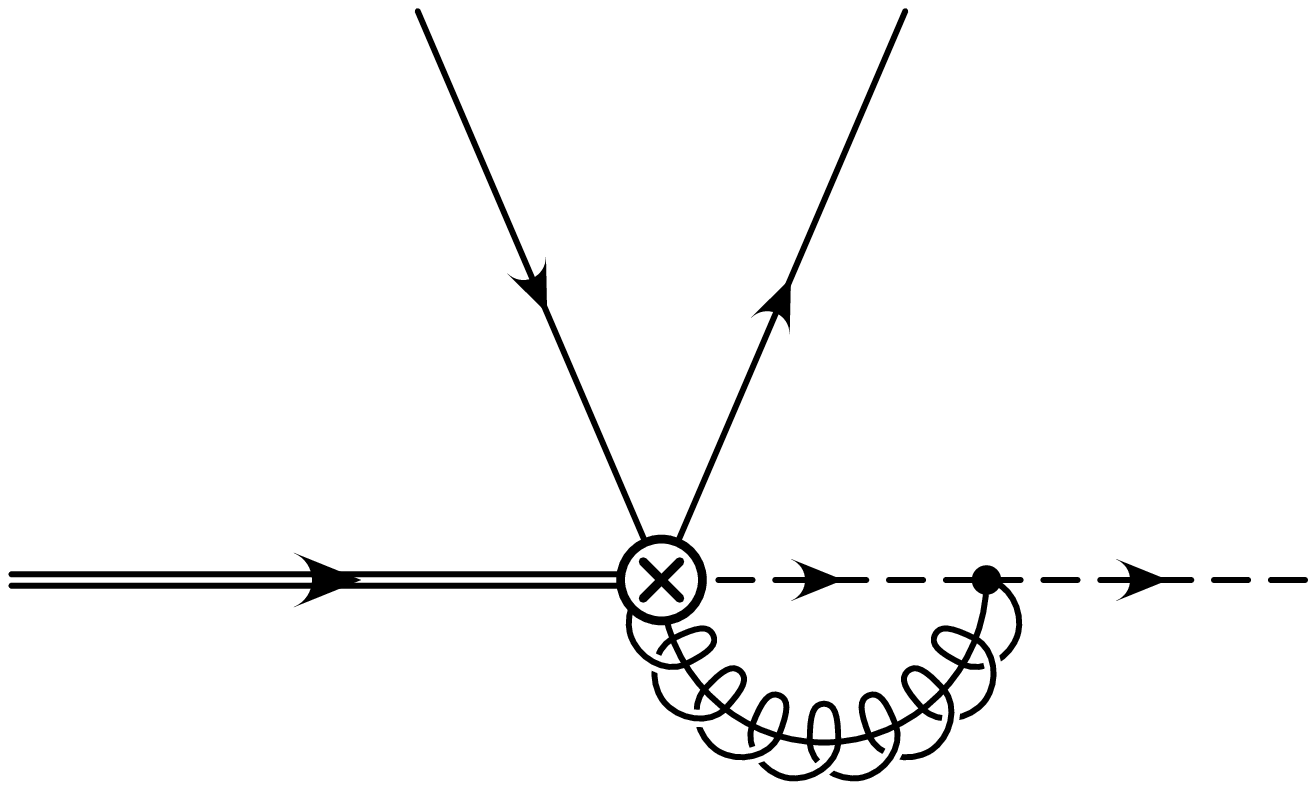}
 \end{minipage}
\caption[The potential gluon exchange between the quarks of $\cc$-pair and collinear gluon emission from the electro-weak vertex.]{The potential gluon exchange between the quarks of $\cc$-pair and collinear gluon emission from the electro-weak vertex. \label{fig:pot}}
\end{center}
\end{figure}

Six diagrams where the ultrasoft gluon is exchanged between the quarks are shown in Fig.~\ref{fig:usoft}. 
Two more diagrams in Fig.~\ref{fig:pot} correspond to the potential gluon exchange between the quarks of the $(\cc)$-pair and the emission of collinear gluon from the electro-weak vertex. 

\begin{figure}[htb]
\begin{center}
\begin{minipage}[h]{5cm}
  \includegraphics[height=1.7cm]{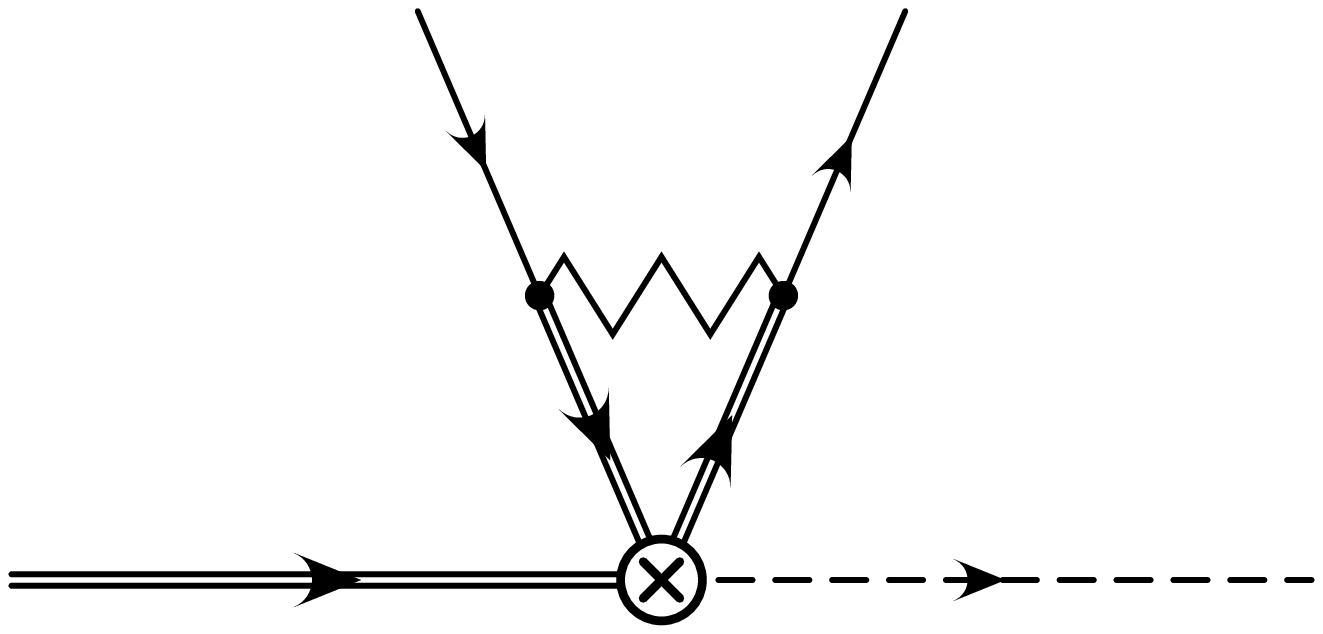}
\end{minipage} 
\begin{minipage}[h]{5cm}
  \includegraphics[height=1.7cm]{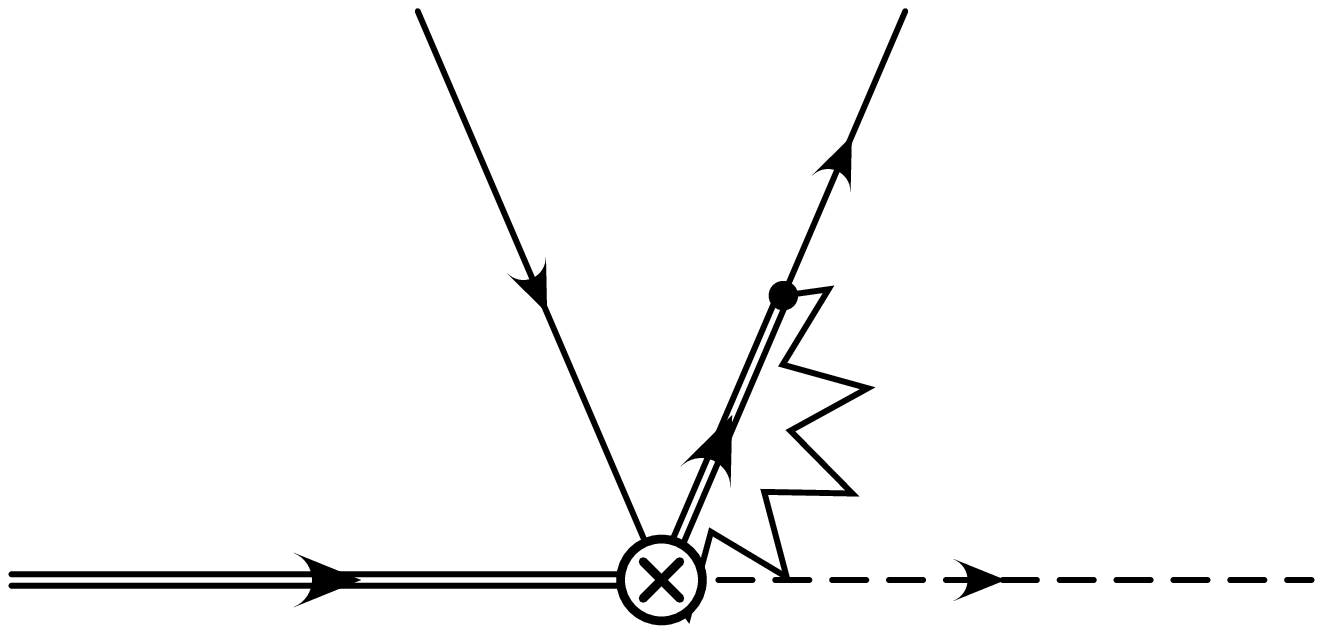}
\end{minipage}
\begin{minipage}[h]{5cm}
  \includegraphics[height=1.7cm]{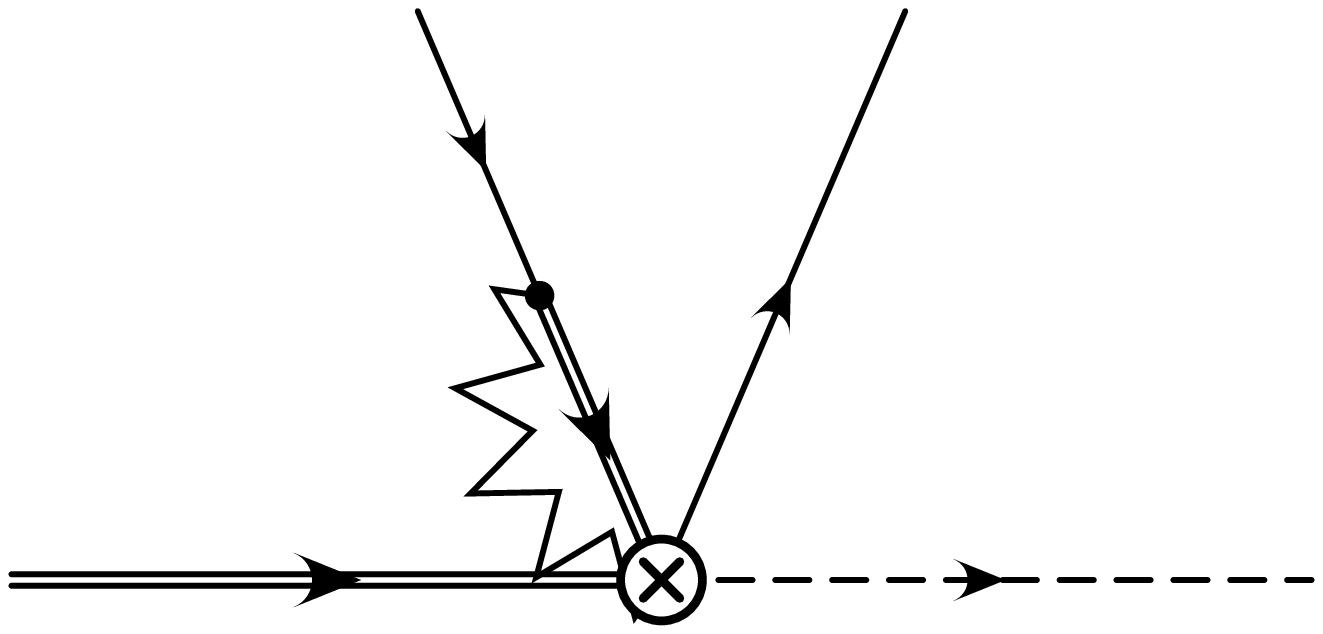}
\end{minipage}
\caption[Soft gluon exchange between the quarks of $\cc$-pair and soft gluon emission from the electro-weak vertex.]{Soft gluon exchange between the quarks of $\cc$-pair and soft gluon emission from the electro-weak vertex. The zigzag line corresponds to soft gluon. The double lines for $c$ and $\bar{c}$-quarks stand for the soft propagators. These diagrams are identically zero in the dimensional regularization.\label{fig:soft}}
\end{center}
\end{figure}

The diagrams involving the soft CNRQCD gluons emitted from the electro-weak vertex are shown in Fig.~\ref{fig:soft}. All these diagrams include also the off-shell soft quarks. The propagators for the soft quarks are HQET propagators because the term $\beta_{\cc}\cdot q_s\sim mv$ and the kinetic energy term in the CNRQCD propagator $p_\perp^2/2m\sim mv^2$ must be discarded. This makes Feynman integrals corresponding to the diagrams in Fig.~\ref{fig:soft} scaleless and thereby identically zero in the dimensional regularization. So the soft gluons do not contribute to the amplitude at one loop. 

The non-renormalized amplitude is given by the sum of the diagrams in Figs.~\ref{fig:usoft} and~\ref{fig:pot}. The ultraviolet (UV) divergences are then removed by the field renormalization factors for the heavy and light quarks, $Z_H$ and $Z_l$, respectively, and by the renormalization matrix of the electroweak operators for $i,j = 0,8$, $Z_{ij}$:
\beqn
&&Z_H=1+\frac{\aS}{4\pi}2C_F\frac{1}{\eps_{UV}},\qquad Z_l=1-\frac{\aS}{4\pi}C_F\frac{1}{\eps_{UV}},\qquad C_F=\frac{4}{3},\nn
&& Z_{00}=1-\frac{\alpha_s}{4\pi}C_F\left\{
  \frac{1}{\eps^2_{UV}}+\frac{2}{\eps_{UV}}\ln\frac{\mu}{2E_s}
    +\frac{5}{2}\frac{1}{\eps_{UV}}\right\},\qquad
     Z_{08}=0,\qquad Z_{80}=0,\nn
    &&Z_{88}=1-\frac{\alpha_s}{4\pi}\left\{
    C_F\left[\frac{1}{\eps^2_{UV}}+\frac{2}{\eps_{UV}}\ln\frac{\mu}{2E_s}\right]
    +\frac{6}{\eps_{UV}}\frac{m_b}{2E_s}\ln\frac{2m_c}{m_b}
    +\frac{3\pi i}{\eps_{UV}}
    +\frac{19}{3}\frac{1}{\eps_{UV}}\right\},
    \label{tech:1}
\eeqn
Here $E_s=\frac{m_b^2-4m_c^2}{2m_b}$ is the energy of the $s$-quark in the $\bRF$. Finally the $\overline{\text{MS}}$-renormalized amplitude is multiplied by the Lehmann-Symanzik-Zimmermann (LSZ)-factor $\sqrt{R}$. The latter is computed from the residues at the poles of heavy and light quark propagators also renormalized in $\overline{\text{MS}}$-scheme. This gives the on-shell matrix element:
\beq
\label{tech:2}
\sqrt{R}= \sqrt{R_H^3\cdot R_l}=1-\frac{\aS}{4\pi}\frac{5}{2}C_F\frac{1}{\eps_{IR}}.
\eeq
The resulting expression for the amplitude is given by:
\beq
\sum_{i=0,8} c_i\sum_{j=1,2,3}[\bar{\xi}_{n,p}\Gamma_j {\bf C}_i\, h_{\beta_b}]
   [\bar{\xi}_{\beta_{\cc}\,\pp}\Gamma_j{\bf C}_i\,\eta^C_{\beta_{\cc}\,-\pp}],\quad {\rm where}
   \label{tech:4}
\eeq
\beqn
&& c_{0}=1+\frac{\alpha_s}{4\pi}C_F\left\{
 - \frac{1}{\eps^2_{IR}}-\frac{2}{\eps_{IR}}\ln\frac{\mu}{2E_s}
    -\frac{5}{2}\frac{1}{\eps_{IR}}+\frac{2\pi i}{v}\left[-\frac{1}{\eps_{IR}}-\ln\frac{\mu^2}{m_c^2v^2}-i\pi\right]\right\},\nn
    &&c_8=1+\frac{\alpha_s}{4\pi}\Big\{
    C_F\left[-\frac{1}{\eps^2_{IR}}-\frac{2}{\eps_{IR}}\ln\frac{\mu}{2E_s}\right]
    -\frac{6}{\eps_{IR}}\frac{m_b}{2E_s}\ln\frac{2m_c}{m_b}
    -\frac{3\pi i}{\eps_{IR}}
    -\frac{19}{3}\frac{1}{\eps_{IR}}\nn
    &&\qquad-\frac{\pi i}{3v}\left[-\frac{1}{\eps_{IR}}-\ln\frac{\mu^2}{m_c^2v^2}-i\pi\right] 
    \Big\}.
    \label{tech:3}
\eeqn 

At one loop there is no mixing between the singlet and the octet operators: $Z_{08}=Z_{80}=0$. One can see why by writing down the color structure of the effective theory diagrams. The diagrams of the first column in Fig.~\ref{fig:usoft} and the diagrams in Fig.~\ref{fig:pot} do not alter the color structure of the vertex. The remaining four diagrams in Fig.~\ref{fig:usoft} actually mix the singlet and octet operators but the interaction vertices for the $c$ and $\bar{c}$ quarks have different signs as it follows from the Feynman rules of the effective theory and therefore the color structure of the sum of the second and the third and/or the fifth and the sixth diagrams in Fig.~\ref{fig:usoft} is either
\beq
T^a{\bf C}\otimes [T^a, {\bf C}]\qquad {\rm or}\qquad {\bf C}T^a\otimes [T^a, {\bf C}].
\eeq
For ${\bf C}=\One$ the commutator vanishes and for ${\bf C}=T^a$ the direct product is reduced to $T^a\otimes T^a$.
%
\subsubsection{Full theory}

\begin{figure}[htb]
\begin{center}
\begin{minipage}[h]{5cm}
  \includegraphics[height=1.5cm]{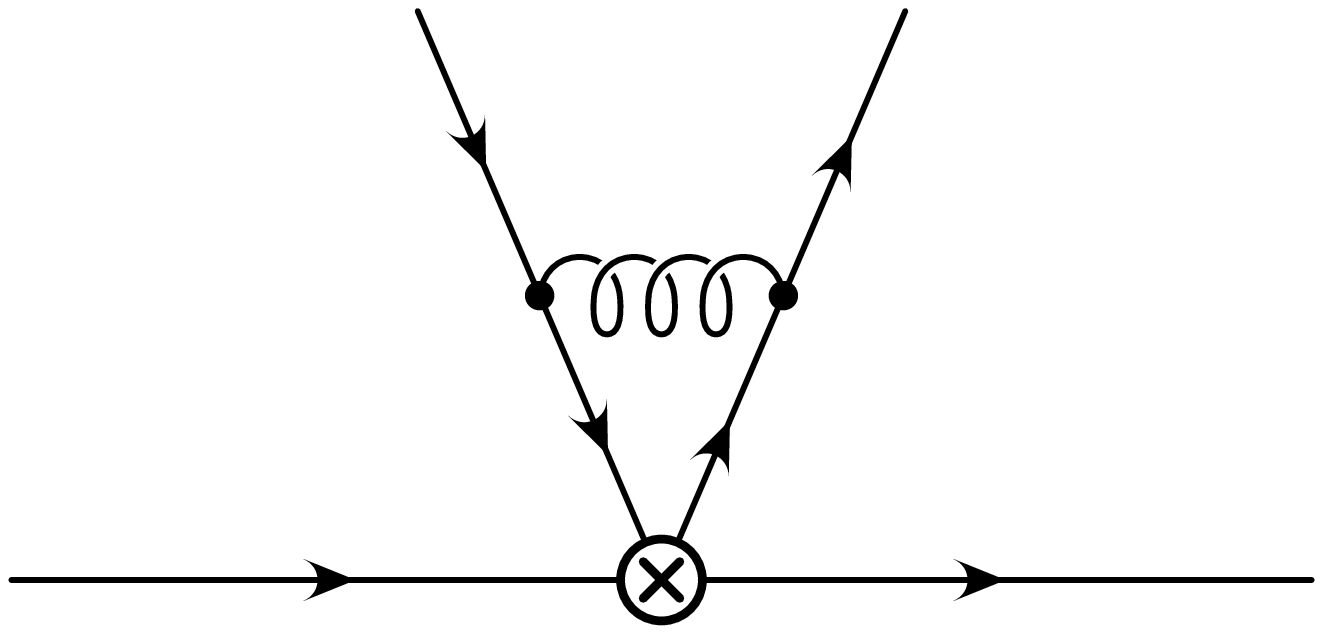}
\end{minipage} 
\begin{minipage}[h]{5cm}
  \includegraphics[height=1.5cm]{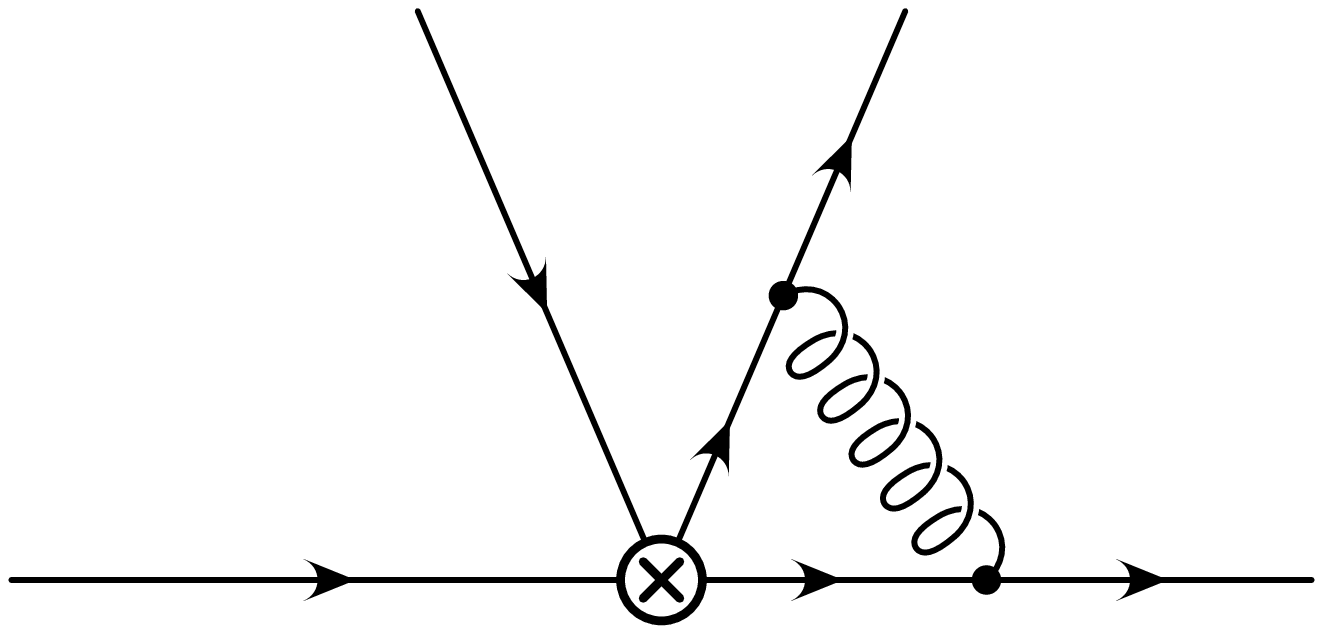}
\end{minipage}
\begin{minipage}[h]{5cm}
  \includegraphics[height=1.5cm]{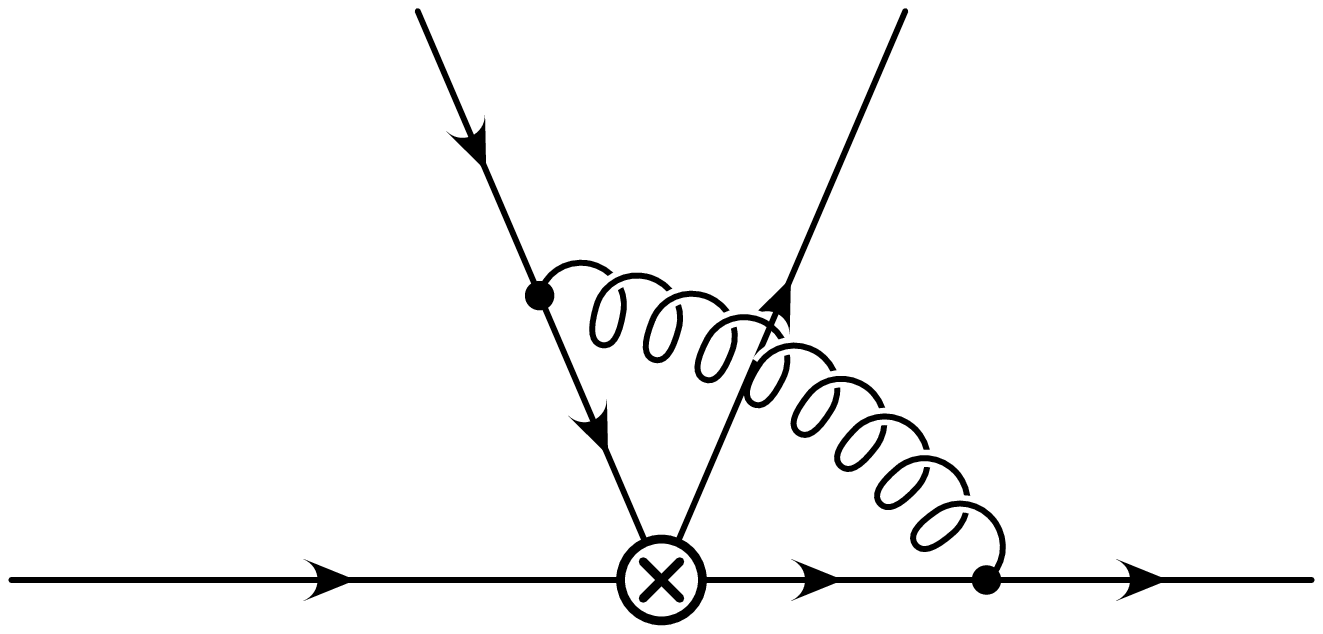}
\end{minipage}
\begin{minipage}[h]{5cm}
  \includegraphics[height=1.7cm]{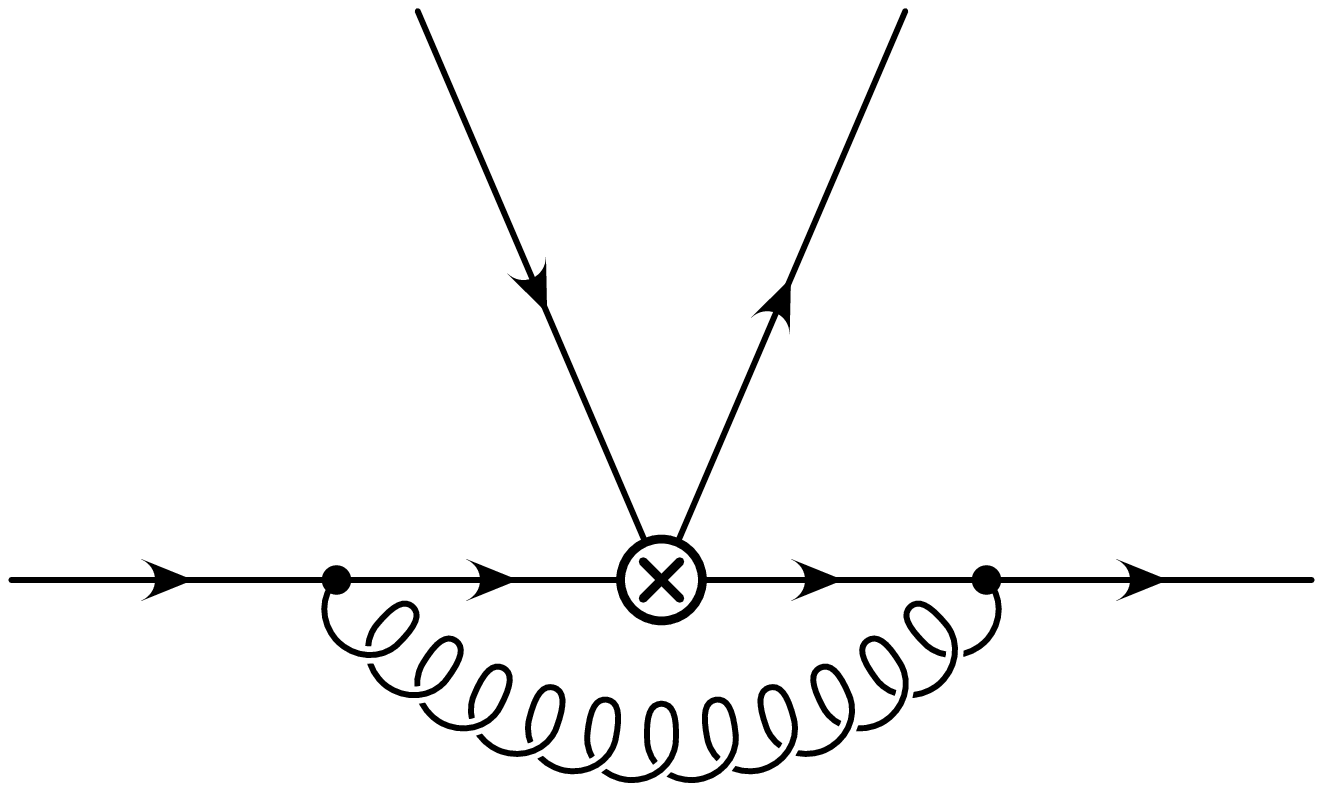}
\end{minipage}
\begin{minipage}[h]{5cm}
  \includegraphics[height=1.5cm]{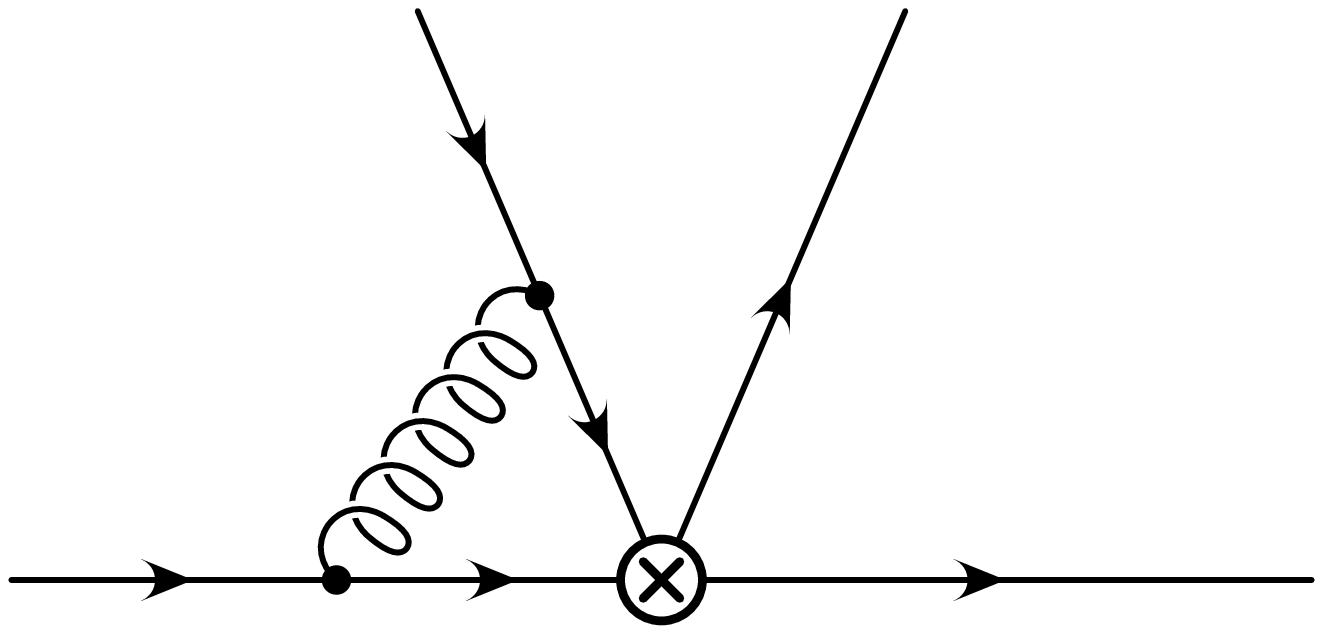}
\end{minipage} 
\begin{minipage}[h]{5cm}
  \includegraphics[height=1.5cm]{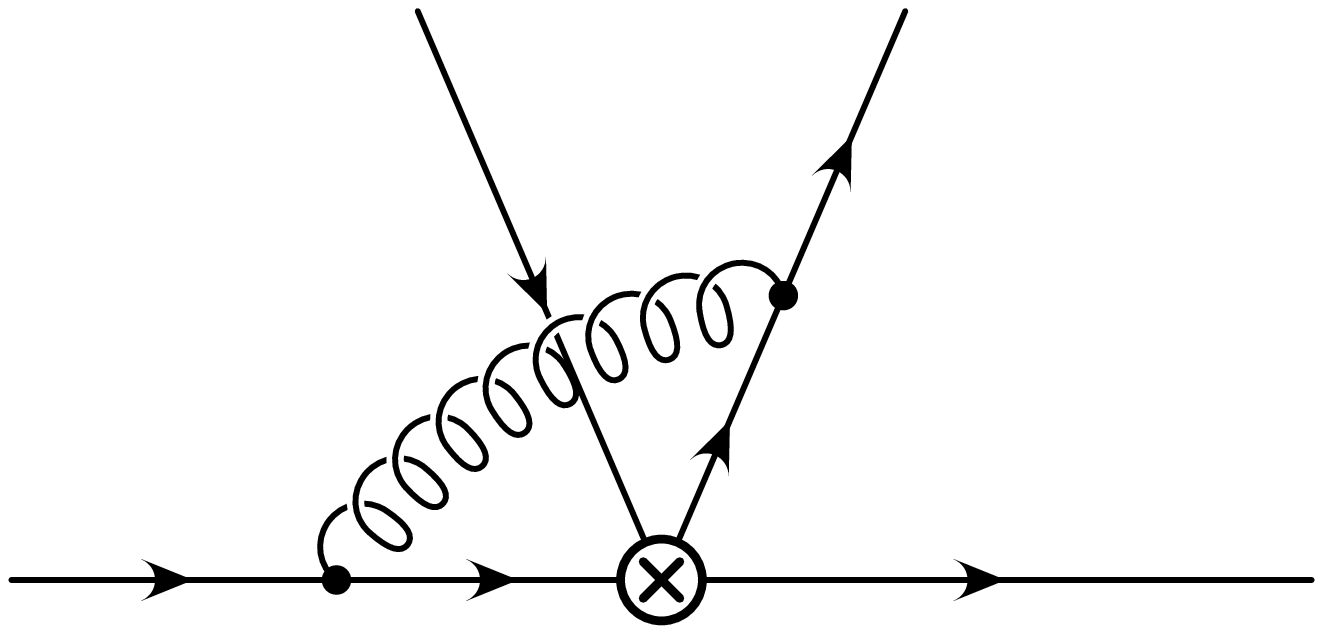}
\end{minipage}
\end{center}
  \caption[FT diagrams contributing to the amplitude.]{\label{fig:FT}Diagrams contributing to the amplitude $b\rightarrow(\cc)\,s$ in the full theory.}
\end{figure}

In the full theory the nonrenormalized amplitude is given by the sum of six diagrams in Fig.~\ref{fig:FT}. 
We are using the NDR scheme for $\gamma_5$. The on-shell value of the full theory amplitude follows when the sum of the six diagrams is multiplied by the field renormalization factors $Z_q$, the operator renormalization matrix $C_i^{\rm bare} = C_j Z_{ji}$, and the LSZ factor:
\beqn
&&Z_q=1-\frac{\alpha_s}{4\pi}C_F\frac{1}{\eps_{UV}},\quad  
Z_{ij} = \delta_{ij} + \frac{\alpha_s}{4\pi}
\frac{1}{\varepsilon_{\rm UV}}
          \left( \begin{array}{cc} 0 & 6 \\ \frac{4}{3} & -2 \end{array}
\right),\nonumber\\[2ex]
&&\sqrt{R_bR_c^2R_s}=1+\frac{\alpha_s}{4\pi}\left(-\frac{10}{3\eps_{UV}}-2\ln\frac{\mu^2}{m_b^2}-4\ln\frac{\mu^2}{m_c^2}-8\right).
\eeqn

The final step is projecting the full theory Dirac structures to the Dirac structures of the effective theory which corresponds to taking the limit $v\rightarrow0$ and $\lambda\rightarrow0$. This is done by means of the reduction formulae listed in Appendix~\ref{sec:10}. The projected full theory amplitude contains only the Dirac structures of Eq.~(\ref{ETstruct1}).
%
\subsubsection{Matching\label{sec:4b}}

At one loop the matrix element on the full theory side upon expanding the Dirac structures in terms of the  effective theory structures becomes:
\beq
	\label{wils:7}
[A^{(0)}_i+\frac{\alpha_s}{4\pi}A^{(1)}_i]\langle O_i\rangle.
\eeq
Here the index $i$ is cumulative and labels both the Dirac and color structure. On the effective theory side the matrix element has the form: 
\beq
	\label{wils:8}
	\left[C^{(0)}_i+\frac{\alpha_s}{4\pi}C^{(1)}_i\right]
	\left[\langle O_i\rangle+\frac{\alpha_s}{4\pi}X^{(1)}_{i\,j} \langle O_j\rangle\right],
\eeq
where $C^{(1)}_i$ is the first order correction to the Wilson coefficient and the second bracket is the matrix element 
in the effective theory at $\order{\aS}$ as it comes from the one-loop diagrams. Equating~(\ref{wils:7}) and~(\ref{wils:8}) gives the matching condition at one-loop:
\beq
	\label{wils:9}
C^{(0)}_i=A^{(0)}_i\quad{\text{and}}\quad	C^{(1)}_i=A^{(1)}_i-C^{(0)}_j\,X^{(1)}_{j\,i}=A^{(1)}_i-A^{(0)}_j\,X^{(1)}_{j\,i}.
\eeq
The coefficients $C^{(0)}_i$ have been introduced in the tree-level matching formula~(\ref{ETstruct1}):
\beq
C^{(0)}_i(r)\in \left\{\frac{1}{2},\,-\frac{1}{4r}\,, -\frac{1}{2}\right\}\quad{\rm with}\quad
r=\frac{m_c}{m_b}. 
\label{wils:2}
\eeq
The analytic expressions of the coefficients $C^{(1)}_i$ are given in Appendix~\ref{sec:12}. Their numerical values at the matching scale of the decay are discussed in section~\ref{sec:6b}. 

The coefficients $X^{(1)}_{j\,i}$ are given by Eq.~(\ref{tech:3}) and contain the IR divergences of the effective theory. The coefficients $A^{(1)}$ contain both the finite part and the IR divergences of the full theory. We confirm the cancellation of the IR-divergences between the full and effective theory amplitudes. From this fact we infer that at the leading order in the effective theory power expansion and in the leading logarithm approximation the effective degrees of freedom include only those listed in section~\ref{fields}. This is the main result of the paper.
%
\subsection{Summing the leading logarithms\label{ewet2all}}

The renormalization matrix $Z_{ij}$ in Eq.~(\ref{tech:1}) is diagonal and contains the terms $\sim1/\eps^2$, $\sim(1/\eps)\log\mu$, and $\sim1/\eps$. The terms proportional to $1/\eps^2$ and $\sim(1/\eps)\log\mu$ give rise to the leading Sudakov logarithms $\sim \alpha_s^n \ln^{n+1}(\mu)$ as we will see shortly. The terms $\sim1/\eps$ give rise to the next-to-leading logarithms $\sim \alpha_s^n \ln^n(\mu)$ and should be combined with the next-to-leading logarithms coming from $\sim 1/\eps^2$ terms in the renormalization matrix $Z_{ij}$ calculated at two loops (see Table~II in~\cite{Bauer:2000yr} for the counting scheme of the divergent terms).  

Discarding the terms $\sim1/\eps$ gives the renormalization factors $Z_{00}=Z_{88}$ which turn out to be equal to the renormalization factor for the $(\bar{s}b)$-current (see~\cite{Bauer:2000yr}):
\beq
	\label{renCon}
    Z_{00}=Z_{88}=1-\frac{\alpha_s}{4\pi}
C_F\left[\frac{1}{\eps^2_{UV}}+\frac{2}{\eps_{UV}}\ln\frac{\mu}{2E_s}\right].
\eeq
Therefore the anomalous dimension matrix (ADM) in the leading logarithm approximation is the same for all operators $\Op_{i\,j}$ and coincides with the ADM of the $(\bar{s}b)$-current in SCET whose Wilson coefficient is calculated in~\cite{Bauer:2000yr}:
\beq
\label{eft:56}
C^{LL}_j\left(\frac{2E_s}{\mu},\,r\right)=C^{(0)}_j(r)\exp\left(\frac{2 C_F}{\beta_0}
\sum_{n=1}^{\infty}\frac{(-1)^n}{n+1}\left[\frac{\beta_0}{2\pi}\right]^n
\alpha_s^n(2E_s)\ln^{n+1}\left[\frac{\mu}{2E_s}\right]\right).
\eeq
We have chosen the form (\ref{eft:56}) for the Wilson coefficient to make it explicit that solving the RG-equation is indeed equivalent to summing up the IR double Sudakov logarithms.

SCET gauge invariance of the jet field $[\bar{\xi}_{n,p}\,W]$ ensures that loop corrections to this operator and to $\bar{\xi}_{n,p}$ are the same~\cite{Bauer:2001ct}. This observation allows one to combine the tree-level matching formula~(\ref{EWETmatch2}) with the Wilson coefficient~(\ref{eft:56}) by treating the latter as the operator function of $\bar{n}\cdot\cP^\dagger$ inserted on the right to the jet field~\cite{Bauer:2001ct}:
\beq
\cL_W^{LL}(\mu,\mu_b)= -\frac{4 G_F}{\sqrt{2}} V_{cb} V^\ast_{cs}
   \sum_{i=0,8}\sum_{j=1}^3 C_i(\mu_b)\,
 [\bar{\xi}_{n,p}\,W\,C^{LL}_j\left(\frac{\bar{n}\cdot\cP^\dagger}{\mu},\,r\right)\Gamma_j {\bf C}_i\,h_{\beta_b}]
   [\bar{\xi}_{\beta_{\cc}\,\pp}\,\Gamma_j{\bf C}_i\,\eta^C_{\beta_{\cc}\,-\pp}].
	 \label{wils:1}
\eeq
Here $\cP^\dagger$ picks up the net label momentum of the collinear operators on the left and $C_i(\mu_b)$ are Wilson coefficients~(\ref{eft:7}). The operator $\bar{n}\cdot\cP^\dagger$ picks up the total momentum of the jet at each order in the $g_s$ expansion. At tree-level this amounts to replacing $2E_s\to\bar{n}\cdot\cP^\dagger$ but when evaluating $\aS$-corrections the presence of the operator function in~(\ref{wils:1}) becomes nontrivial. The dependence of $\cL_W$ on the scales $\mu$ and $\mu_b$ is to remind that the Lagrangian is obtained in two steps: firstly by running from the electroweak scale to $\mu_b=2E_s$ and then to $\mu$. 

%
%
\section{Factorization\label{sec:5}}

Applying the EWET Lagrangian~(\ref{wils:1}) to the decay of the $B$-meson requires evaluating the matrix elements of the Lagrangian between the hadronic states of the effective theory. In this section we prove that in the leading logarithm approximation the matrix element of the EWET Lagrangian between the effective theory states factorizes into the product of the matrix elements of the heavy-to-light current and decay constant of the charmonium state. 
%
\subsection{Ultrasoft-free EWET Lagrangian}

As we have seen the ultrasoft gluons are the only degrees of freedom common to all three sectors of the effective theory in the leading logarithm approximation. The following field redefinition (see e.g.~\cite{pa:0109045}) makes it possible to rewrite the EWET Lagrangian~(\ref{wils:1}) in terms of the ultrasoft-free field operators by incorporating the ultrasoft gluons into Wilson lines:
\beqn
\label{usdec}
&&\xi_{n,p}=Y_n\,\xi_{n,p}^{(0)}, \qquad A_{n,q}=Y_n\,A_{n,q}^{(0)}Y_n^\dagger,\qquad
W=Y_n\, W^{(0)}\,Y_n^\dagger,\nn
&&h_{\beta_b}=Y_{\beta_b}h_{\beta_b}^{(0)},\nn
&&\xi_{\beta_{\cc}\,\pp}=Y_{\beta_{\cc}}\xi_{\beta_{\cc},\pp}^{(0)},
\qquad 
\eta^C_{\beta_{\cc}\,-\pp}=Y_{\beta_{\cc}}\eta^{C\,(0)}_{\beta_{\cc},-\pp}.
\eeqn
Here $Y_l$, with $l=n,\beta_b$, or $\beta_{\cc}$, stands for the path ordered exponent:
\beq
Y_l(x)=P\exp\left(i g_s \int^x_{-\infty}ds\, l\cdot A_{us}(l\,s)\right).
\eeq
It is a well-known fact that ultrasoft gluons do not renormalize the four-quark Coulomb interaction term in NRQCD (see e.g.~\cite{pa:9910209}) and so, as long as soft gluons are ignored, the ultrasoft-free CNRQCD Lagrangian includes only the ultrasoft-free potential quarks and potential gluons. 

Applying these field redefinitions to~(\ref{wils:1}) gives:
 \beq
 \label{der:22}
\Op^{LL}_{ij}=[\bar{\xi}_{n,p}^{(0)}W^{(0)} \,C^{LL}_j\left(\frac{\bar{n}\cdot\cP^\dagger}{\mu},\,r\right)\,Y_n^\dagger \Gamma_j {\bf C}_i \,Y_{\beta_b}\,\,h_{\beta_b}^{(0)}]
 [\bar{\xi}_{\beta_{\cc}\,\pp}^{(0)}\, Y_{\beta_{\cc}}^\dagger \Gamma_j {\bf C}_i Y_{\beta_{\cc}}\,\eta^{C\,(0)}_{\beta_{\cc}\,-\pp}].
\eeq
It is instructive to pause and check that at one loop corrections to~(\ref{der:22}) exactly reproduce the effective theory amplitude given by~(\ref{tech:4}) and~(\ref{tech:3}). Expanding the ultrasoft and collinear Wilson lines in~(\ref{der:22}) to the first order in $g_s$ reproduces the diagrams shown in Fig.~\ref{fig:usoft} and the second diagram in Fig.~\ref{fig:pot}. Terms of the order $g_s^2$ coming from expanding the $Y_l(x)$ are self-energy diagrams and must be discarded since only the amputated diagrams contribute according to LSZ prescription. Wilson lines $Y_l(x)$ bring the renormalization factors which reproduce the renormalization factors of the quark fields due to the ultrasoft self-energy diagrams. Expanding $Y_{\beta_{\cc}}$ reproduces only the HQET part of the potential quark propagator in Fig.~\ref{FRules}. This is sufficient because the corresponding diagrams (2nd, 3rd, 5th, and 6th diagrams in Fig.~\ref{fig:usoft}) do not have $1/v$-divergences. Physically for the ultrasoft gluon the CNRQCD quark is the same as HQET quark as long as the relative motion of the quarks in the $\ccCMF$ is ignored. 
%
\subsection{Matrix element for $B\to \psi X_s$.}

Consider the matrix element of~(\ref{der:22}) between the states $\brac{{X_s}_n,\, \psi_{\beta_{\cc}}}$ and $\ket{0_{\beta_{\cc}},\,B_{\beta_b}}$, where ${X_s}_n$ is any state which includes a collinear $s$-quark. The states $\ket{0_{\beta_{\cc}}}$ and $\brac{\psi_{\beta_{\cc}}}$ do not contain ultrasoft gluons, so the matrix element of the singlet operator factorizes:
\beq
\label{eft:33}
\brac{{X_s}_n}
[\bar{\xi}_{n,p}^{(0)}W^{(0)}\,C^{LL}_j\left(\frac{\bar{n}\cdot\cP^\dagger}{\mu},\,r\right)\,\Gamma_j Y_n^\dagger \, Y_{\beta_b}\,\,h_{\beta_b}^{(0)}]
\ket{B_{\beta_b}}
\brac{\psi_{\beta_{\cc}}}
[\bar{\xi}_{\beta_{\cc}\,\pp}^{(0)}\Gamma_j \,\eta^{C\,(0)}_{\beta_{\cc}\,-\pp}]
\ket{0_{\beta_{\cc}}}.
\eeq

Now let us show that the matrix element of the octet operator (${\bf C}=T^a$) vanishes and therefore the color state of the $\cc$-pair does not contribute to the decay rate (in the leading logarithm approximation). To this end it would suffice to show vanishing of the correlator
\beqn
\label{eft:31}
&&\brac{{X_s}_n,\,0_{\beta_{\cc}}}\text{T}\Big\{ \Op_{\cc}^{(0)}(x)\,
[\bar{\xi}_{n,p}^{(0)}W^{(0)}\,C^{LL}_j\left(\frac{\bar{n}\cdot\cP^\dagger}{\mu},\,r\right)\,\Gamma_j \,Y_n^\dagger\, T^a\, Y_{\beta_b}\,\,h_{\beta_b}^{(0)}] \nn
&&~~~~~~~~~~~~\times[\bar{\xi}_{\beta_{\cc}\,\pp}^{(0)}\Gamma_j \,
Y_{\beta_{\cc}}^\dagger T^a Y_{\beta_{\cc}}\, \eta^{C\,(0)}_{\beta_{\cc}\,-\pp}](y)
\Big\}\ket{0_{\beta_{\cc}},\,B_{\beta_b}},
 \eeqn
where $\Op_{\cc}^{(0)}(x)$ is the ultrasoft-free interpolating operator carrying quantum numbers of $\psi_{\beta_{\cc}}$, the bound state  of the $\cc$-pair. The matrix elements~(\ref{eft:31}) and~(\ref{eft:33}) with ${\bf C}_i=T^a$ are related by LSZ reduction formula and vanishing~(\ref{eft:31}) implies that~(\ref{eft:33}) is also zero. 

\begin{figure}[h]
\begin{center}
\includegraphics[height=5cm]{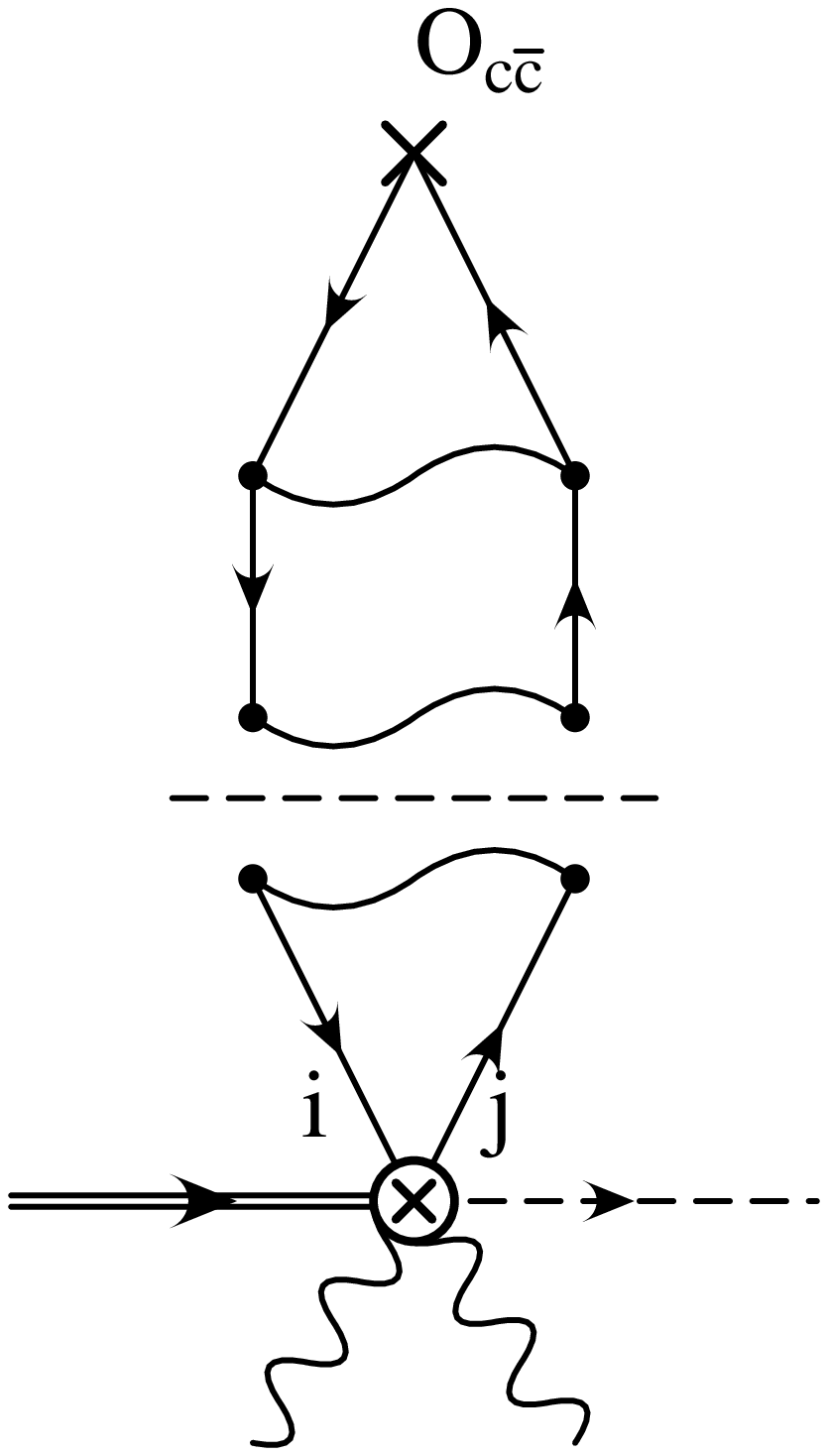}
\end{center}
\caption[The ladder diagrams that contribute to the T-product]{\label{ladder}  Ladder diagrams that contribute to the T-product~(\ref{eft:31}) at the leading order in $\alpha_s$. Wiggly lines represent ultrasoft gluons which couple between themselves and to the external state $B$. The collinear gluons (not shown) couple to the collinear $s$-quark line and to the collinear degrees of freedom in $X_s$.}
\end{figure}

The diagrams contributing to the T-product in~(\ref{eft:31}) are the ladder diagrams shown in Fig.~\ref{ladder}, where the potential gluons are exchanged between the quarks of $\cc$-pair. 
The interpolating operator $\Op_{\cc}^{(0)}(x)$ and the currents in~(\ref{eft:31}) are color singlets, therefore the color structure of the diagram in Fig.~\ref{ladder} is:
\beq
\label{eft:32}
\sim A^{a,\,a_1,a_2\ldots} \tr \left\{T^{b_1}\,T^{b_2}\ldots [\ldots [T^a,\,T^{a_1}],\,T^{a_2}],\ldots ]\ldots T^{b_2}\,T^{b_1}\right\}
\eeq
Here $a_i$ and $b_i$ are adjoint indices. The commutators inside the trace arise upon expansion of the ultrasoft operators $Y_{\beta_{\cc}}$. The insertions of $T^{b_i}$ on both sides are due to potential gluon exchange. The quantity $A^{a,\,a_1,a_2}$ represents the color structure arising when contracting the ultrasoft fields from the $\cc$-current with the fields of the heavy-to-light current, and with the ultrasoft degrees of freedom in the $B_{\beta_b}$ state. Expanding the commutator inside the trace gives the structure proportional to $T^a$. Contracting the indices $b_i$ on both sides of $T^a$ again gives a structure proportional to $T^a$. Therefore the whole trace is proportional to the trace of $T^a$ and vanishes.

%
%
\section{Beyond the leading logarithms\label{sec:6}}

If it were not for the soft NRQCD gluons next-to-leading corrections could be included systematically into $C_{i\,j}(\bar{n}\cdot \cP^\dagger/\mu,r)$ of Lagrangian~(\ref{wils:1}). Without the soft gluons we simply have three HQET quarks, two of them propagate along the same Wilson line in the opposite directions and as the matching shows decouple from the $(\bar{s}b)$ current. To see explicitly whether soft gluons violate factorization in the next-to-leading logarithm approximation two loop calculation in the effective theory is required. We have not attempted this calculation, although made the first step in this direction by performing the tree-level matching for the EWET Lagrangian that includes both collinear and soft NRQCD gluons.
%
\subsection{Including soft gluons}

The matching is performed using the technique explained in the Appendix A of~\cite{pa:0109045}. The method is to introduce the auxiliary fields corresponding to the off-shell quarks and gluons, to write down the Lagrangian for the auxiliary  and on-shell fields, and then to integrate out the auxiliary fields by solving  the equations of motion (EOM). The integration corresponds to the tree-level matching to all orders in $g_s$. The result given by~(\ref{EWETA3c}) does not exactly reproduce the corresponding result for the heavy-to-light current in SCET (see e.g. Eq.~(A22) in~\cite{pa:0109045}) because the soft NRQCD gluons are not protected by the SCET symmetry under the soft gauge transformations. In this section we briefly outline the derivation.

\subsubsection{Auxiliary fields}

\begin{figure}[h]
\begin{center} 
  \includegraphics[height=2cm]{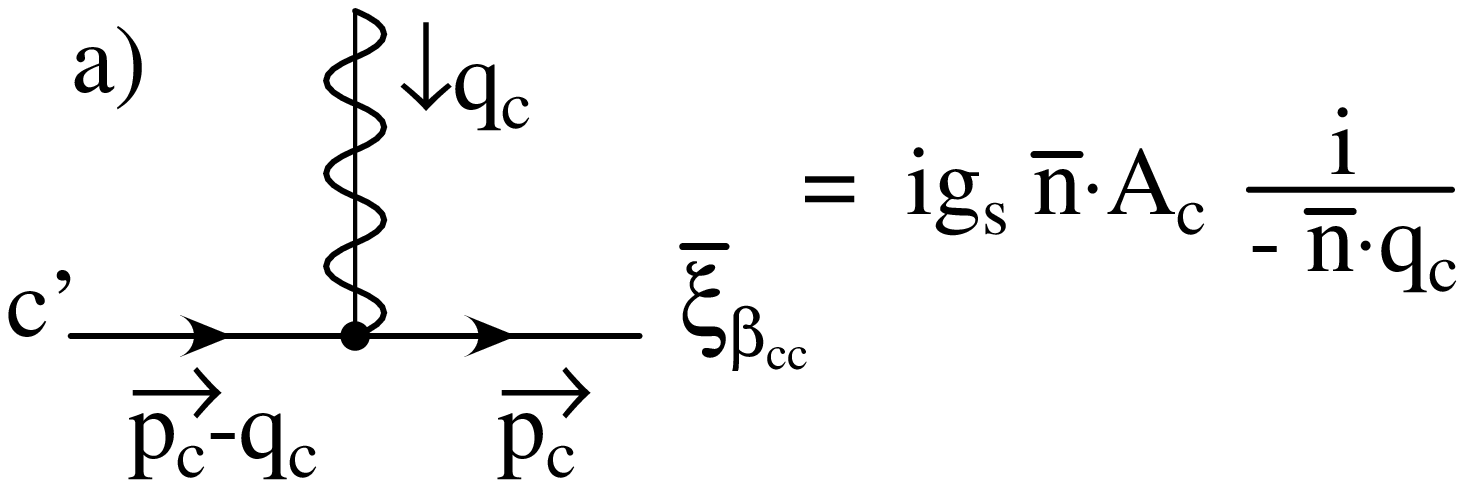}
  \includegraphics[height=2cm]{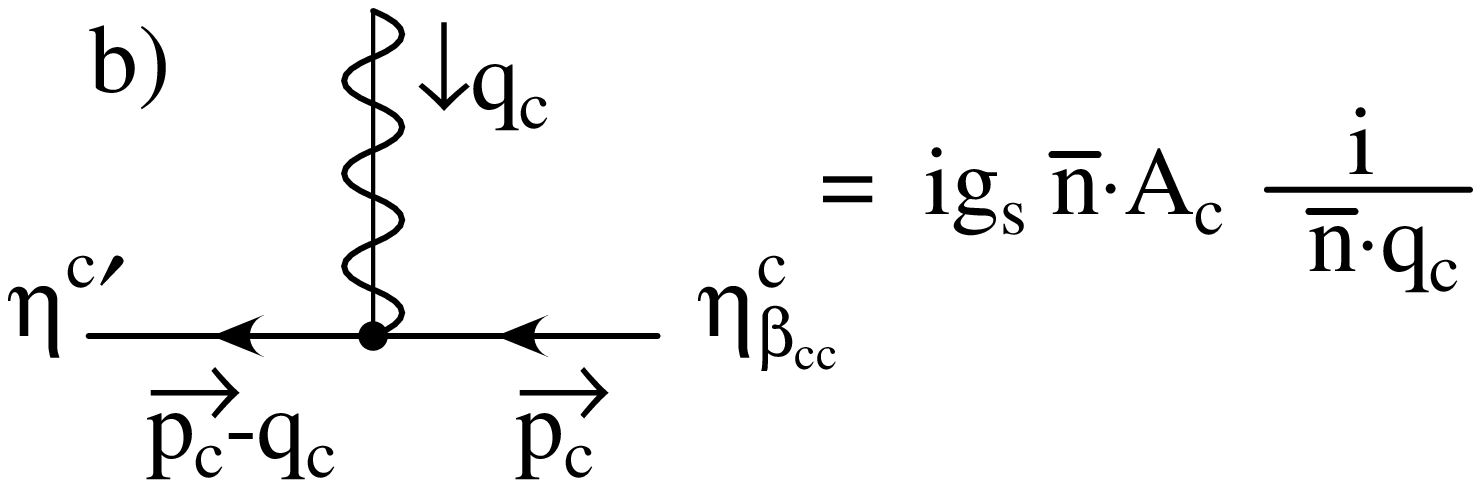}
\end{center}
  \caption[Collinear gluon interacting with $\eta^c$-field of the antiquark and the off-shell mode ${\eta^c}'$.]{a) Collinear gluon interacting with $c$-quark and the off-shell mode $c'$. Interaction between collinear gluon and  the off-shell mode $c'$ is given by the same vertex. b) Collinear gluon interacting with $\eta^c$-field of the antiquark and the off-shell mode ${\eta^c}'$. Interaction between collinear gluon and the off-shell mode ${\eta^c}'$ is given by the same vertex. The overall sign is opposite relative to~a) due to the reversed fermion flow. \label{aux2}}
\end{figure}

\begin{figure}[h]
\begin{center}
  \includegraphics[height=2cm]{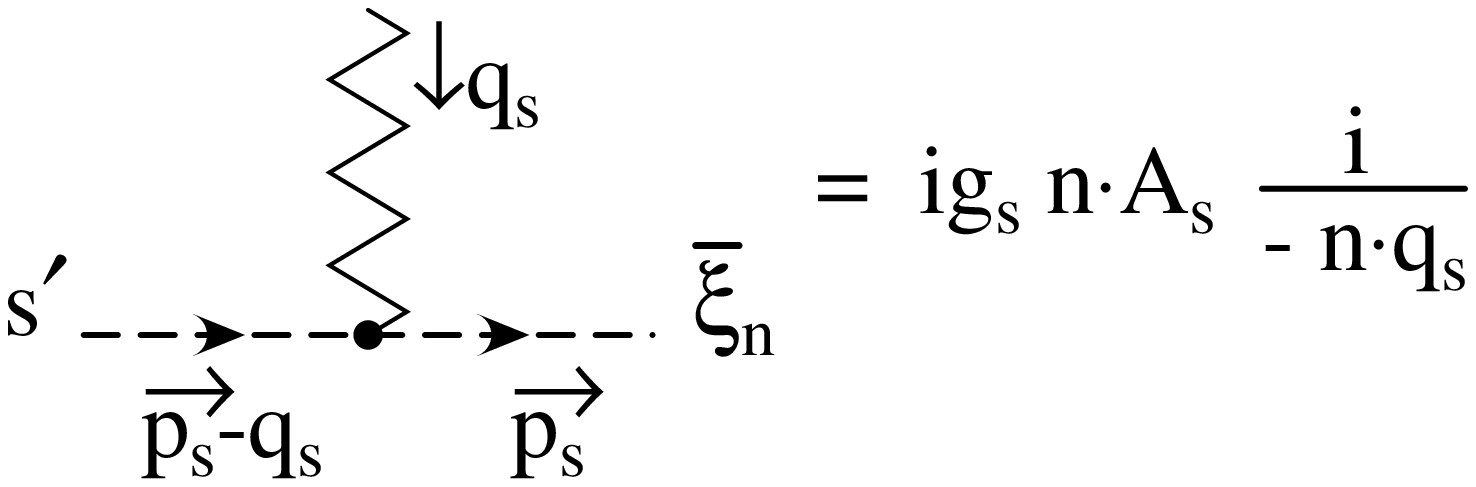}
\end{center}
  \caption[Soft gluon interacting with $s$-quark and the off-shell $s'$-field.]{Soft gluon interacting with $s$-quark and the off-shell $s'$-field. Interaction between collinear gluon and the off-shell mode $s'$ is given by the same vertex.\label{aux5}}
\end{figure}

The first step is to draw all tree-level diagrams that introduce the off-shell fields. Figures~\ref{aux2} and~\ref{aux5} show how the off-shell modes arise for the quarks of $\cc$-pair and $s$-quark, respectively.

\begin{figure}[h]
\begin{center}
  \includegraphics[height=1.8cm]{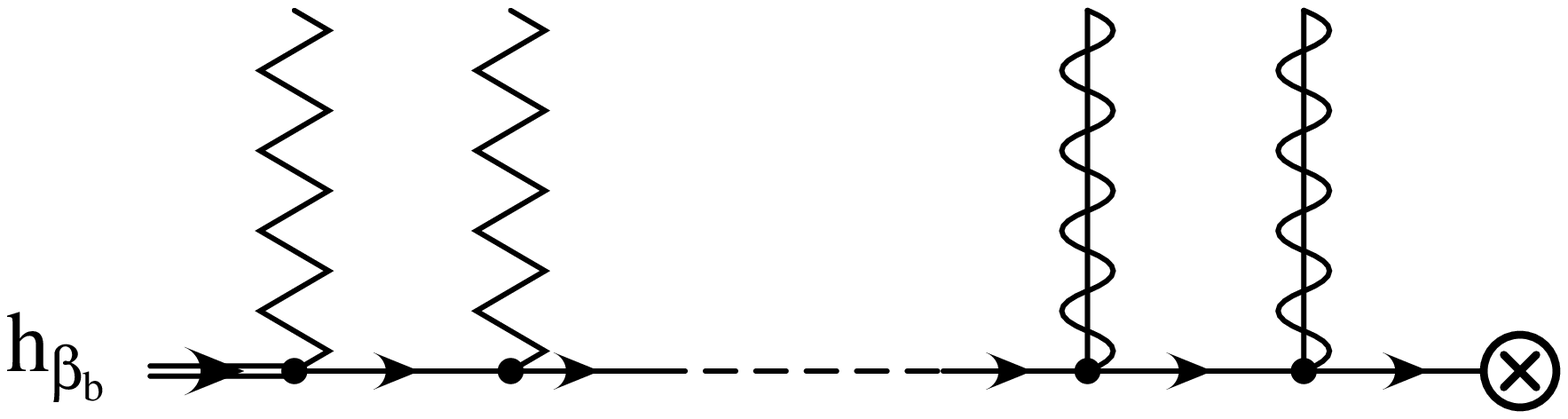}
\end{center}  
\caption[Diagrams which contribute to the tree-level matching of the heavy $b$-quark field at the LO in the ET power expansion.]{Diagrams which contribute to the tree-level matching of the heavy $b$-quark. Soft gluons (zigzag lines) must be next to the heavy quark field.\label{bmatch}}
\end{figure}

\begin{figure}[h]
\begin{center}
  \includegraphics[height=2cm]{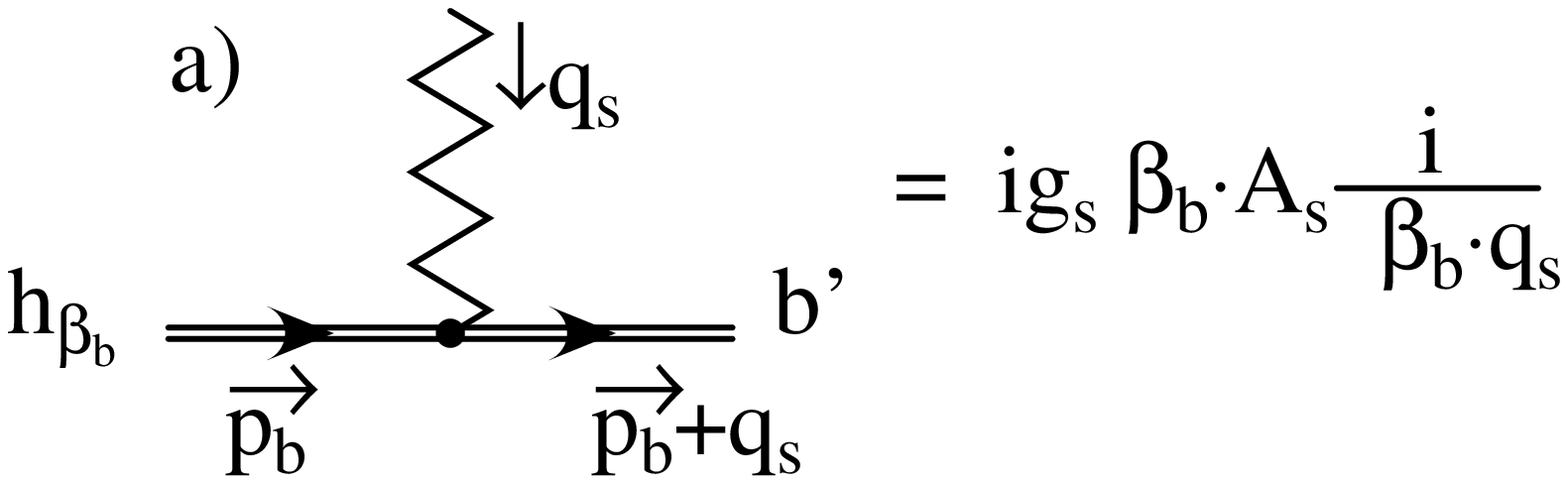}
  \includegraphics[height=2cm]{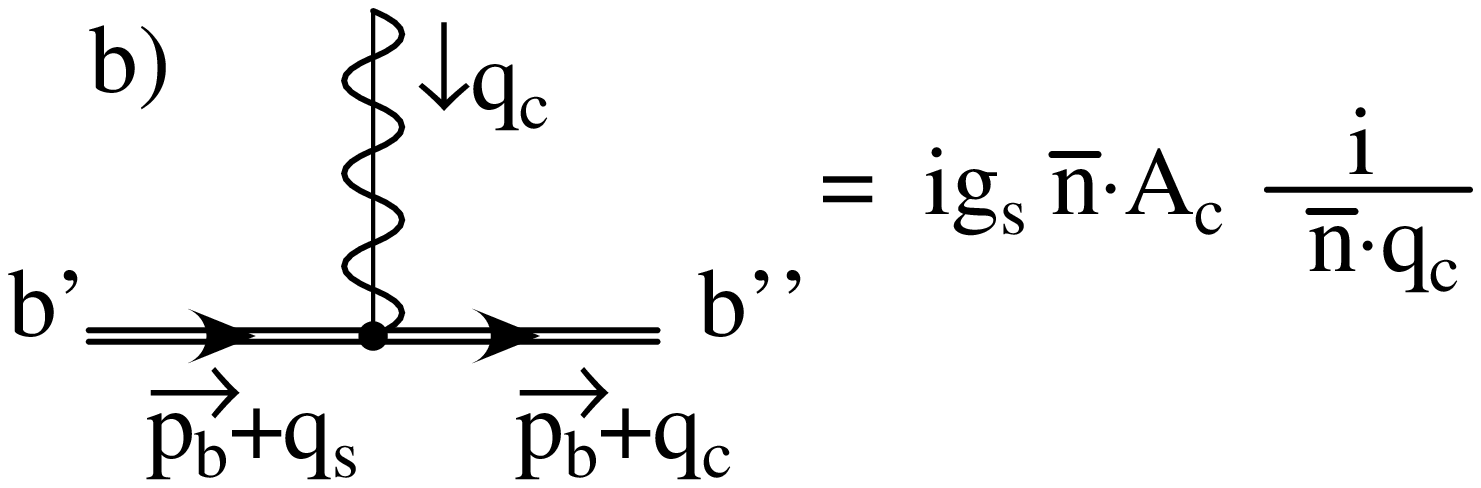}
\end{center}
  \caption[Collinear gluon interacting with the off-shell mode $b'$ and off-shell mode $b''$.]{a) Soft gluon interacting with the $b$-quark and the off-shell mode $b'$. The interaction between the soft gluon and the off-shell mode $b'$ is given by the same vertex. b) Collinear gluon interacting with the off-shell mode $b'$ and the off-shell mode $b''$. The interaction between the collinear gluon and the off-shell mode $b''$ is given by the same vertex.\label{aux3}}
\end{figure}

For the $b$-quark we have to match in two steps. Both soft and collinear gluons give the $b$-quark an off-shell momentum but the corresponding off-shellnesses are of different orders of magnitude. A soft momentum is $\sim mv$ and collinear momentum is $ \sim m$. Once the collinear momentum is transferred to the $b$-quark any off-shell momentum brought by the soft gluons is subleading compared to the collinear momentum propagating along the line. A diagram where a soft gluon is inserted between collinear gluon(s) and the electro-weak vertex contributes the factor
\beq
\frac{\beta_b\cdot A_s}{(\beta_b\cdot n)(\bar{n}\cdot q_c)}\sim \order{v},
\eeq
where $q_c$ is the collinear momentum propagating along the $b$-quark line. (A gauge field scales like the momentum it transfers.) Therefore only the diagrams with the soft gluons next to the heavy quark field $h_{\beta_b}$ and the collinear gluons next to the electro-weak vertex shown in Fig.~\ref{bmatch} will contribute. 
To account for this we introduce two off-shell fields for the $b$-quark: $b'$ which is off-shell by $\sim mv$ and $b''$ which is off-shell by $\sim m$, (see Fig.~\ref{aux3}).

\begin{figure}[h]
\begin{center}
  \includegraphics[height=2.3cm]{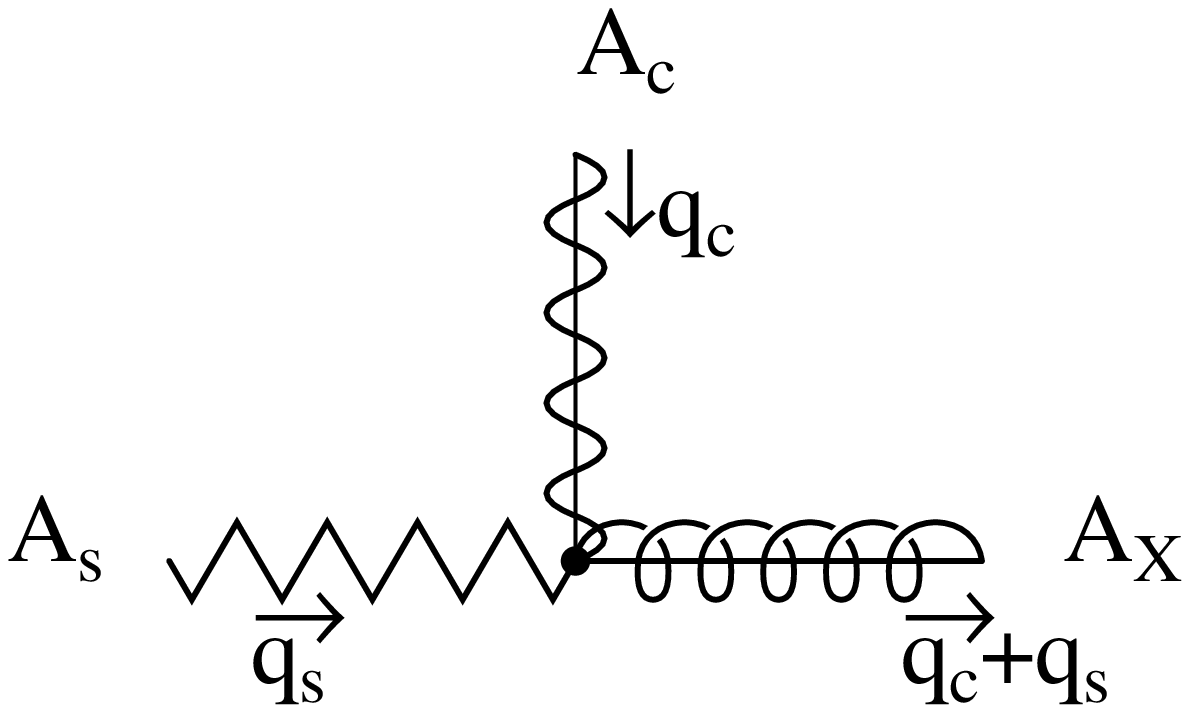}
 \caption[Soft gluon fusing with a collinear gluon into the off-shell field.]{Soft gluon fusing with a collinear gluon into the off-shell field  $A_X$.\label{aux6}}
 \end{center}
\end{figure}

In addition to the soft and collinear gluons we should take into account the mode arising when a soft gluon fuses with a collinear one, see Fig.~\ref{aux6}. The auxiliary gluon field $A_X$ transfers momentum $q_X\sim m (1,v,v)$ in the light-cone coordinates (compare to the collinear momentum $q_c\sim(1,\lambda,\lambda^2)$). Recall that $v\sim\lambda$ for the decay kinematics. At the leading order in the effective theory power expansion the interaction vertex of the field $A_X$ with the heavy quarks is the same as for collinear gluons. The interaction vertex of $A_X$ and the $s$-quark is the same as for a soft gluon. The gluon Lagrangian for $A_X$ is discussed in~\cite{pa:0109045}. 
\subsubsection{Matching}

Writing down the auxiliary Lagrangians which generate the Feynman rules listed above is straightforward. The EWET operator is given by:
\beq
\label{EWETmatch3}
\Op^{(\text{tree})}=
    [(\bar{\xi}_{n,p}+\bar{s}')\Gamma_j {\bf C}\, (b''+b')]
   [(\bar{\xi}_{\beta_{\cc}\,\pp}+\bar{c}')\Gamma_j{\bf C}\,(\eta^c{}'+\eta^C_{\beta_{\cc}\,-\pp})].
\eeq 
Here the auxiliary fields are understood as perturbative expansions of solutions for the corresponding EOM following from the auxiliary Lagrangians. For the $b$-quark we have to solve first the EOM for the $b'$ field and then use the solution to solve EOM for $b''$. The solutions are  
\beqn
\label{solstuff}
&&\bar{c}'=\bar{\xi}_{\beta_{\cc},\pp}(W_X^\dagger-1),\nn
&&\eta^c{}'=(W_X-1)\eta^c_{\beta_{\cc},-\pp},\nn
&&\bar{s}'=\bar{\xi}_{n,p}(S_{n,\,X}^\dagger-1),\nn
&&b''=(W_X-1)b',\nn
&&b'=S_{\beta_b}\,h_{\beta_b},
\eeqn
where $S_{\beta_b}$, $W_X$, and $S_X$ are Wilson lines
\beqn
\label{wxsx}
&&S_{\beta_b}=\sum_{\text{perm}}\exp\left(\frac{1}{\beta_b\cdot \cP}\,g_s \beta_b\cdot A_{s,\,q}\right),\nn
&&W_X=\sum_{\text{perm}}\exp\left(-\frac{1}{\bar{n}\cdot \cP} g_s [\bar{n}\cdot A_{X,\,q}+\bar{n}\cdot A_{c,\,q}]\right),\nn
&&S_{n,\,X}^\dagger=\sum_{\text{perm}}\exp\left( g_s [n\cdot A_{X,\,q}+n\cdot A_{s,\,q}]\frac{1}{n\cdot \cP^\dagger}\right).
\eeqn
Combining Eqs.~(\ref{EWETmatch3}),~(\ref{solstuff}), and~(\ref{wxsx}) gives:
 \beq
\label{EWETmatch}
\Op^{(\text{tree})}=
    [\bar{\xi}_{n,p}\,S_X^\dagger\,W_X\Gamma_j {\bf C}\,\, S_{\beta_b}\,h_{\beta_b}]
   [\bar{\xi}_{\beta_{\cc}\,\pp}\,\Gamma_j{\bf C}\,\eta^C_{\beta_{\cc}\,-\pp}].
\eeq

Solving the EOM for the auxiliary field $A_X$ makes it possible to write $\Op^{(\text{tree})}$ in terms of $A_c$ and $A_s$ only. This remarkable calculation is performed in Appendix A of~\cite{pa:0109045}, to which the reader is referred. The only difference is that here we work with the soft NRQCD gluons instead of the soft SCET gluons. The result is:
\beq
\label{ansatz}
S^\dagger_X\,W_X=W\,S_n^\dagger,
\eeq
where $W$ is the collinear Wilson line~(\ref{wl}) and $S_n$ is given by:
\beq
\label{ws}
S_n^\dagger=\sum_{\text{perm}}\exp\left( g_s n\cdot A_{s,\,q}\frac{1}{n\cdot \cP^\dagger}\right).
\eeq
Using~(\ref{ansatz}) eliminates the auxiliary field $A_X$ from~(\ref{EWETmatch}) which gives:
\beq
\label{EWETA3b}
\Op^{(\text{tree})}= [\bar{\xi}_{n,p}W\,S^\dagger_n\,\Gamma_j {\bf C}\,S_{\beta_b}\,h_{\beta_b}]
  [\bar{\xi}_{\beta_{\cc}\,\pp}\Gamma_j \,{\bf C}\, \eta^C_{\beta_{\cc}\,-\pp}].
\eeq
It is easy to verify this formula at $\order{g_s}$ by explicit matching.

Now the net momentum of the $\cc$-pair is transversal with respect to the 4-velocity $\beta_{\cc}$.  Therefore the net momentum $q_s$ of the soft gluons emitted from the EWET vertex must be transversal as well because they end up inside the $(\cc)$-current: $\beta_{\cc}\cdot q_s=0$. This observation suggests that~(\ref{EWETA3b}) must be written as
\beq
\label{EWETA3c}
\Op^{(\text{tree})}= [\bar{\xi}_{n,p}W\,\delta_{\beta_{\cc}\cdot\cP}\,S^\dagger_n\,\Gamma_j {\bf C}\,S_{\beta_b}\,h_{\beta_b}]
  [\bar{\xi}_{\beta_{\cc}\,\pp}\,\Gamma_j\,{\bf C}\,\eta^C_{\beta_{\cc}\,-\pp}],
\eeq
where the Kronecker delta enforces transversality of the net soft momentum transferred from $\bar{s}b$- to $\cc$-current. Equation~(\ref{EWETA3c}) is the final result.
\subsubsection{A remark}

Equation~(\ref{EWETA3c}) does not hint that the factorization holds in the next-to-leading logarithms approximation and beyond. However, the transversality constraint results in something resembling the factorization. Consider Coulomb gauge. In this gauge gluons are transversal, so the soft gluons obey: $\beta_{\cc}\cdot A_s=0$. Now let us apply the transversality constraint in~(\ref{EWETA3c}) {\it separately} to the soft momenta emitted from the $b$- and $s$-quark lines. (Actually the constraint restricts the {\it net} momentum.) Then the Wilson lines $S_n$ and $S_{\beta_b}$  become equal. To see this notice that at the leading order in power expansion the momentum conservation reads:
\beq
\beta_b=\frac{2m_c}{m_b}\beta_{\cc}+\frac{E_s}{m_b}n.
\eeq
Applying this equation to the argument of $S_{\beta_b}$ (\ref{wxsx}) together with the transversality constraints yields:
\beq
\frac{\beta_b\cdot A_{s,\,q}}{\beta_b\cdot \cP}=\frac{n\cdot A_{s,\,q}}{n\cdot \cP}\qquad\rightarrow\qquad S_n=S_{\beta_b}.
\eeq

The last equation allows one to move the soft gluons from the $(\bar{s}b)$ current to $(\cc)$- current, so the factorization ``holds'' even when the soft NRQCD gluons are included. However the attempts to make anything rigorous out of the above observation were unsuccessful. Most probably because the factorization is broken beyond the leading logarithm approximation.

%
\subsection{Wilson coefficients in the next-to-leading logarithm approximation at $\mu=m_b$\label{sec:6b}}

The one-loop matching (see section~\ref{sec:4b}) gives the initial values of the Wilson coefficients $C^{(1)}_i$ at the matching scale~$\mu=\mu_b$. At this scale the EWET Lagrangian in the next-to-leading logarithm approximation is:
\beqn
       \label{wils:4}
&& \cL_W^{NLL}(\mu_b,\mu_b)= -\frac{4 G_F}{\sqrt{2}} V_{cb} V^\ast_{cs}
   \sum_{i,k=0,8}\sum_{j=1}^3 C_i(\mu_b)\Big\{C^{(0)}_j(r)\delta_{i\,k}+\frac{\alpha_s(\mu_b)}{4\pi}
C^{(1)}_{i\,k;j}(\frac{m_b}{\mu_b},\,r)\Big\}\Op^{(0)}_{k\,j}.
\eeqn
The numerical values of $C^{(1)}_{i\,k;j}$ for $\mu_b=m_b$ and $r=m_c/m_b=0.30$ are: 
\beqn
&&C^{(1)}_{i\,k;1}(1,\,0.30)=\left(
\begin{array}{cc}
-11.27 & -11.27 - 3.41 i   \\
 - 2.50 - 0.76 i & 8.94 - 11.55 i
\end{array}
\right),\nn
&&C^{(1)}_{i\,k;2}(1,\,0.30)=\left(
\begin{array}{cc}
 16.11 & 19.64 + 6.58 i\\
 4.36+1.46 i & -13.56 +20.29 i
\end{array}
\right),\nn
&&C^{(1)}_{i\,k;3}(1,\,0.30)=\left(
\begin{array}{cc}
-10.75 &  10.81 +1.23 i\\
2.40+0.27 i & -7.87+9.00 i
\end{array}
\right).
\label{wils:5}
\eeqn 
Here indices $i$ and $k$ label the rows and columns, respectively.

Comparatively large numerical values of the non-diagonal elements $i\neq k$ of the coefficients $C^{(1)}_{i\,k;j}$ together with the observation that numerically the Wilson coefficient $C^{NLO}_8(m_b)$ is about ten times larger than the Wilson coefficient of the singlet (see section~\ref{sec:2}) suggest that the next to the leading logarithm correction to~(\ref{wils:1}) is of the same order of magnitude. 
%
%
\section{Conclusion\label{sec:7}}

In this paper we have proposed the effective theory for the decay $B\to \psi + h$, where $\psi$ is a
charmonium state and $h$ a light hadron neglecting the effects due to the spectator quark. We have identified the relevant degrees of freedom and verified explicitly by doing the matching calculation at one loop that the effective theory reproduces all the IR divergences of the perturbative QCD. We have also shown that in the leading logarithm approximation the effective theory decay amplitude factorizes into the decay constant of the charmonium state and the matrix element of heavy-to-light current. The latter has been extensively studied. 

However including into consideration the soft NRQCD gluons most probably ruins the factorization beyond the leading logarithms. At least the tree-level matching result~(\ref{EWETA3c}) does not suggest the opposite. The comparatively large value of $\alpha_s\sim v\sim 0.6$ in charmonium implies that the higher order corrections cannot be ignored. To achieve a better quantitative understanding of non-leptonic $B$-decays into charmonium we should definitely go beyond the leading logarithms. Our results can be used e.g. for calculating the RG-improved Lagrangian in the next-to-leading logarithm approximation provided the anomalous dimension matrix of the effective theory is known to two loops. 

\vspace{5mm}

M.S. is grateful to prof. S.~Fleming for numerous conversations during the work on the project and to prof. A.~Manohar for discussing the results.

The work of C.B.~was supported by the Bundesministerium f\"ur Bildung und Forschung, Berlin-Bonn. 
The work of B.G. and M.S was supported in part by the US  Department of Energy under contract DE-FG03-97ER40546.
\appendix
%
%
\section{Semi-inclusive decay $B\to J/\psi +h$.\label{sec:11}}

In this section we apply the EWET Lagrangian~(\ref{wils:1}) to the calculation of the semi-inclusive decay rate of $B\to J/\psi +h$. Due to factorization of the charmonium state in the leading logarithm approximation the decay can be treated along the same lines as the semi-leptonic decay $B\to  X_s\gamma$~\cite{Bauer:2001yt}.
The derivation below is presented in some detail. The result is given by~(\ref{ddratefinal2}).
\subsection{Kinematics}

The differential decay rate is given by
\beq
\label{ddrate2}
\frac{d\Gamma}{dE_{J/\psi}}=\frac{|{\bf p}_{J/\psi}|}{(2\pi)^2}\sum_{X_s,\,\eps}\frac{|\brac{J/\psi,\,X_s}\cH_W\ket{B}|^2}{2M_B}
(2\pi)^4\delta^{4}(p_B-p_{J/\psi}-p_X).
\eeq
Here $|{\bf p}_{J/\psi}|=\sqrt{E_{J/\psi}^2-M_{J/\psi}^2}$. The differential decay rate depends only on $E_{J/\psi}$ that varies in the range
\beq
\label{EJrange}
E_{J/\psi}\in\left[M_{J/\psi},\,\frac{M_B^2+M_{J/\psi}^2}{2M_B}\right]\approx[3.1,\,3.56]\,\text{GeV}.
\eeq
There is a region of $E_{J/\psi}$ where our effective theory can be used. In this region the state $X_s$ must be collinear, i.e. the SCET parameter $\lambda=M_X/2E_X$, where $M_X$ is the invariant mass of the jet, is small. On the other hand $M_X$ must be reasonably larger than $\LamConf$, so the jet is inclusive.

It is customary to take $\lambda\sim\sqrt{\LamConf/E_X}\ll 1$ as a value of SCET parameter for a jet which is both collinear and inclusive. The kinematic expressions for $E_X$ and $M_X$ in terms of $E_{J/\psi}$ are
\beq
\label{ddrate4}
E_X=M_B-E_{J/\psi}\qquad\text{and}\qquad M_X^2=M_B^2+M_{J/\psi}^2-2M_BE_{J/\psi},
\eeq
so we have to look for a region in~(\ref{EJrange}) where 
\beq
\label{condition}
\frac{M_X}{2E_X}\sim\sqrt{\frac{\LamConf}{E_X}}\ll1.
\eeq
For $\LamConf\sim0.35$~GeV such a region exists near the center of the range~(\ref{EJrange}) where $\lambda\sim0.4$ and $M_X\sim1.6$~GeV. The SCET parameter is reasonably small, on a par with $v\sim0.6$ in the charmonium.
%
\subsection{$J/\psi$ decay constant}

The matrix element in~(\ref{ddrate2}) factorizes according to~(\ref{eft:33}). So, the first step is the evaluation of the matrix element of the $(\cc)$-current (see~(\ref{ETstruct1})):
\beq
\label{MEJpsi5}
\brac{J/\psi_{\beta_{\cc}}}
[\bar{\xi}_{\beta_{\cc}\,\pp}^{(0)}\Gamma_j \,\eta^{C\,(0)}_{\beta_{\cc}\,-\pp}]
\ket{0_{\beta_{\cc}}},\quad\text{where}\quad
\Gamma_j=\left\{\dslash{n}_\perp,\,\gamma_5,\,\dslash{\eps}_{+\perp}\right\}.
\eeq
The matrix element of $\gamma_5$ vanishes by parity; it is non-zero for the pseudoscalar state $\eta_c$. The other two can be expressed via the $J/\psi$ decay constant. In the full theory it is defined as (see~\cite{pa:bensign}):
\beq
\label{deconstJ}
\brac{J/\psi(p,\,\eps)}[\bar{\psi}(x)\gamma_\mu \psi(x)]\ket{0}=-i f_{J/\psi}M_{J/\psi}\eps_\mu^\ast(p)\,
e^{i p\cdot x}.
\eeq
Here $M_{J/\psi}$ is the mass, and $\eps_\mu^\ast(p)$ is the polarization vector of the state. The decay constant determined from the leptonic decay mode $J/\psi\to l^+l^-$ is $f_{J/\psi}=(405\pm15)$MeV \cite{pa:bensign}. 

To match the left-hand side of~(\ref{deconstJ}) to the CNRQCD at the leading order in $v$-expansion we replace the full theory operators with their expressions in terms of the ultrasoft-free effective theory fields (see~(\ref{operators}) and~(\ref{usdec})). The states of the full theory are replaced with the states of ultrasoft-free CNRQCD. In the sum over $\pp$ and ${\pp}^\prime$ only the terms where $\pp+{\pp}^\prime=0$ survive. Also we should take into account the Wilson coefficient $C_{J/\psi}(m_{red}/\mu)$, where $m_{red}=m_c(\mu)/2$ is the reduced mass of the $\cc$-pair, that comes upon matching between the full and effective theory decay constants (the coefficient $C_0(m_{red}/\mu)$ in~\cite{pa:9501296}). The exponent on the left-hand side is simply $2m_c\beta_{\cc}\cdot x$ since taking into account the binding energy of the $\cc$-pair (see discussion below~(\ref{scale})) would amount to replacing $2m_c\to 2m_c(1-v^2/8)$, which brings corrections of $\order{v^2}$.

On the right-hand side of~(\ref{deconstJ}) we substitute $p_{J/\psi}=2m_c\beta_{\cc}\equiv M_{J/\psi}\beta_{cc}$. Equating both sides gives:
\beq
\label{deconstJET}
\sum_{\pp\neq0}\brac{J/\psi_{\beta_{\cc}}(\eps)}[\bar{\xi}_{\beta_\cc,\,\pp}^{(0)}\,\gamma^\mu\, \eta^{C\,(0)}_{\beta_\cc,-\pp}]\ket{0}=-i C^{-1}_{J/\psi}(m_c/2\mu)f_{J/\psi}\,M_{J/\psi}\,\eps_\mu^\ast\,(\beta_{\cc}).
\eeq
%
\subsection{Heavy-to-light current}

The Hamiltonian density $\cH_W$ in the leading logarithm approximation is given by $-\cL^{LL}_W$ and its matrix element by~(\ref{eft:33}). We write it here explicitly for reference:
\beqn
\label{MEJpsi}
&&\brac{J/\psi_{\beta_{\cc}}(\eps),\,X_s}\cL^{LL}_W(\mu,\,\mu_b)\ket{B_{\beta_b}}=\frac{4 \GF}{\sqrt{2}} V^{}_{cb} V^\ast_{cs}\,
C_0( \mu_b )\sum_{p\neq0}\sum_{p_\perp\neq0}\sum_je^{i(M_{J/\psi}\beta_{\cc}+\cP-m_b\beta_b) x}\nn
&&\times\brac{{X_s}_n}
[\bar{\xi}_{n,p}^{(0)}W^{(0)}\,C^{LL}_j\left(\frac{\bar{n}\cdot\cP^\dagger}{\mu},\,r\right)\,\Gamma_j Y_n^\dagger(x) \,h_{\beta_b}(x)]
\ket{B_{\beta_b}}\brac{J/\psi_{\beta_{\cc}}(\eps)}
[\bar{\xi}_{\beta_{\cc}\,\pp}^{(0)}\Gamma_j \,\eta^{C\,(0)}_{\beta_{\cc}\,-\pp}]
\ket{0_{\beta_{\cc}}}.\nn
\eeqn 
Replacing the matrix element of the $\cc$-current with~(\ref{deconstJET}) gives:
\beqn
\label{MEJpsi2}
&&\brac{J/\psi_{\beta_{\cc}}(\eps),\,X_s}\cL^{LL}_W(\mu,\,\mu_b)\ket{B_{\beta_b}}=\frac{4 \GF}{\sqrt{2}} V^{}_{cb} V^\ast_{cs}\,
C_0( \mu_b )[-i C^{-1}_{J/\psi}(m_c/2\mu) f_{J/\psi}\,M_{J/\psi}]\nn
&&\times \frac{1}{2}\sum_{p\neq0}\left\{ \brac{{X_s}_n}\,
J_{n,\,\beta_b}(P_R;\,x)
[\eps^\ast(\beta_{\cc})\cdot n_\perp] - 
J_{n,\,\beta_b}(\dslash{\eps}_-(n);\,x)
[\eps^\ast(\beta_{\cc})\cdot \eps_{+\perp}(n)]\ket{B_{\beta_b}}\right\},
\eeqn 
where 
\beq
\label{ETcurrent}
J_{n,\,\beta_b}(\Gamma_j;\,x)=e^{i(M_{J/\psi}\beta_{\cc}+\cP-m_b\beta_b)\cdot x}[\bar{\xi}_{n,p}^{(0)}W^{(0)}\,C^{LL}\left(\frac{\bar{n}\cdot\cP^\dagger}{\mu}\right)\,\Gamma_j\, Y_n^\dagger \,h_{\beta_b}](x)
\eeq 
is the standard leading order expression for the SCET heavy-to-light current slightly modified here by the presence of the phase factor of the $\cc$-state. The function $C^{LL}(\bar{n}\cdot\cP^\dagger/\mu)$ represents the exponential factor of~(\ref{eft:56}), the coefficients $C^{(0)}_{1,3}(r)$ are written explicitly (see~(\ref{wils:2})). 

The sum over collinear labels in~(\ref{MEJpsi2}) is approximated by the single term corresponding to the stationary point of the exponent. The label $p$ is chosen so that the momentum components in the directions $n$ and $\perp$ vanish. In other words only oscillations on the ultrasoft scale are permitted in the exponential factor in~(\ref{ETcurrent}). In so doing we have to keep the residual component along $\bar{n}$ which is of the same order $\sim\lambda^2$ as the ultrasoft momentum of the matrix elements in~(\ref{MEJpsi2}). Expanding the exponent in~(\ref{ETcurrent}) in light-cone coordinates in the $B$-meson frame, where $2\beta_b=\bar{n}+n$, gives:
\beq
\label{momcon10}
[M_{J/\psi}(\beta_{\cc}\cdot\bar{n})+\bar{n}\cdot\cP-m_b]\frac{n}{2}+\cP_\perp+[M_{J/\psi}(\beta_{\cc}\cdot n)-m_b]\frac{\bar{n}}{2}.
\eeq
At the stationary point:
\beq
\label{labelcon}
M_{J/\psi}(\beta_{\cc}\cdot\bar{n})+\bar{n}\cdot\cP-m_b=0\qquad\text{and}\qquad\cP_\perp=0.
\eeq
Momentum conservation in the $B$-meson frame reads:
\beq
\label{kinematika}
M_B\beta_b=M_{J/\psi}\beta_{\cc}+\frac{n}{2}(E_X+|\bp_X|)+\frac{\bar{n}}{2}(E_X-|\bp_X|)
\quad\text{and}\quad\bp_{J/\psi}=\bp_X.
\eeq
Using it we solve the first equation in~(\ref{labelcon}) for $\bar{n}\cdot\cP$: 
\beq
\bar{n}\cdot\cP=m_b-M_B+E_X+|\bp_X|=2E_X(1+\order{\lambda^2})
\eeq
and express the residual component along $\bar{n}$ in terms of $E_{J/\psi}$. Finally, taking the sum over collinear labels in~(\ref{MEJpsi}) amounts to replacing the effective theory current~(\ref{ETcurrent}) by the expression
\beq
\label{ETcurrent2}
J_{n,\,\beta_b}(\Gamma_j;\,x)=e^{i(E_{J/\psi}+|{\bf p}_{J/\psi}|-m_b)\frac{1}{2}(\bar{n}\cdot x)}[\bar{\xi}_{n,p}^{(0)}W^{(0)}\,\Gamma_j\, Y_n^\dagger \,h_{\beta_b}](x)
\eeq
and substituting for the operator function $C^{LL}(\bar{n}\cdot\cP^\dagger/\mu)$ its value at the stationary point $C^{LL}(2E_X/\mu)$. Note that although $\cP_\perp$ is set to zero, the $\cP_{\perp}^2$ is the invariant mass of the jet~$M_X^2$ (see~(\ref{ddrate4})) and must be kept.

Taking the square of~(\ref{MEJpsi2}) and summing over polarizations of $J/\psi$ gives:
\beqn
\label{MEJpsi3}
&&|\brac{J/\psi_{\beta_{\cc}}(\eps),\,X_s}\cL^{LL}_W(\mu,\,\mu_b)\ket{B_{\beta_b}}|^2=2 \GF^2\,|V^{}_{cb} V^\ast_{cs}|^2\,
\left[C_0( \mu_b )\,C^{LL}\left(\frac{2E_X}{\mu}\right) C^{-1}_{J/\psi}\left(\frac{m_c}{2\mu}\right)\right]^2\,f_{J/\psi}^2\,\times\nn
&&\times\left\{M_B^2\,| \brac{{X_s}_n}
J_{n,\,\beta_b}(P_R;\,0)
\ket{B_{\beta_b}}|^2+M_{J/\psi}^2|\brac{{X_s}_n}
J_{n,\,\beta_b}(\dslash{\eps}_-(n);\,0)
\ket{B_{\beta_b}}|^2\right\}.
\eeqn 
%
\subsection{Differential decay rate}

Summing over the inclusive collinear states in the last line of~(\ref{MEJpsi3}) is performed by means of the standard trick. The sum is equal to the imaginary part of the T-product of the effective theory currents~(\ref{ETcurrent2}):
\beq
\label{ETJJvsDR}
- 2\, \text{Im} \,T_j(p_{J/\psi})\quad\text{where}\quad 
T_j(q)=-i\int d^{\,4} x \, e^{-i q\cdot x}\frac{1}{2M_B}
\brac{B_{\beta_b}}\text{T}\{J^\dagger_{n,\,\beta_b}(\Gamma_j;\,x)\,J_{n,\,\beta_b}(\Gamma_j;\,0)\}\ket{B_{\beta_b}}.
\eeq
Time-ordering in this expression is applied only to the fields which carry labeled momenta. The ultrasoft fields belong to the zero bin~\cite{pa:zerobin} of the sum over collinear momenta and are not affected by the time ordering, the corresponding operators are simply averaged over the $B_{\beta_b}$-state. 

The matrix element in~(\ref{ETJJvsDR}) is then reduced to a convolution of the perturbative jet-function and a shape function of $B$-meson which is a non-perturbative object. Those functions are defined as:
\beqn
\label{Jetfunction}
&&\brac{0}\text{T}[W^{\dagger\,(0)}\,\xi_{n,p}^{(0)}](x)\,[\bar{\xi}_{n,p^\prime}^{(0)}W^{(0)}]\ket{0}=i\int \frac{d^{\,4}k}{(2\pi)^4}e^{-ikx}\,J_P(k^+)\,\frac{\dslash{n}}{2},\nn
&&S(l^+)=\frac{1}{2}\brac{B_{\beta_b}}[\bar{h}_{\beta_b}\,\delta(i\,n\cdot D_{us}-l^+)\,h_{\beta_b}]\ket{B_{\beta_b}}.
\eeqn
Here $k^+=n\cdot k$ and $l^+=n\cdot l$. The calculation is almost identical to $B\to  X_s\gamma$~\cite{Bauer:2001yt} to which the reader is referred~\footnote{The derivaition given there makes use of the identity
$$
\text{P}\exp(i \int^{1}_{0}ds\,x\, A(x\,s))\,f(0)=\sum_{n=0}^\infty\frac{(-x)^n}{n!}D^n(x)\,f(x)
=\exp[-x\,D(x)]\,f(x),
$$
where $D(x)=\partial-iA(x)$ with $A(x)$ being a (matrix) function of a single variable. The identity is not difficult to prove by showing that both the path-ordered exponent and the series change identically under small variations of $x$.}. The only difference is due to Dirac structures. In our case $\Gamma_j$ is either $P_R$ or $\dslash{\eps}_-(n)$. To average the Dirac structures over $B_{\beta_b}$-meson state we use the fact that the B-meson is a pseudoscalar and in HQET the spin of the heavy quark decouples from the gluon field at the leading order in $\LamConf/m_b$-expansion. The calculation is straightforward and gives:
\beq
\label{PR3eps3}
\brac{B_{\beta_b}}\gamma^0\,P_R^\dagger\,\gamma^0 \,\frac{\dslash{n}}{2}\,P_R\ket{B_{\beta_b}}=\frac{1}{4}\quad\text{and}\quad
\brac{B_{\beta_b}}\gamma^0\,\dslash{\eps}_-(n)^\dagger\,\gamma^0 \,\frac{\dslash{n}}{2}\,\dslash{\eps}_-(n)\ket{B_{\beta_b}}=\frac{1}{2}.
\eeq
 
Finally the differential decay rate in terms of the jet and shape functions becomes:
\beqn
\label{ddratefinal2}
\frac{d\Gamma}{dE_{J/\psi}}&&=
\frac{|{\bf p}_{J/\psi}|}{4\pi}
 \GF^2\,|V^{}_{cb} V^\ast_{cs}|^2\,
\left[C_0( \mu_b )\,C^{LL}\left(\frac{2E_X}{\mu} \right) C^{-1}_{J/\psi}\left(\frac{m_c}{2\mu}\right)\right]^2\,f_{J/\psi}^2\,
\left(M_B+2\frac{M_{J/\psi}^2}{M_B}\right)\nn
&&\times
   \int d\,k^+\,\left[-\frac{1}{\pi}\text{Im}\,J_P(k^+-(E_{J/\psi}+|{\bf p}_{J/\psi}|-m_b),\mu)\right]\, S(k^+,\mu).
\eeqn
The $\mu$-dependence of the jet and shape functions cancels that one of the Wilson coefficients, so the differential decay rate is $\mu$-independent. The argument of the jet function contains the invariant mass of the jet, $\cP^2_\perp\sim M_X^2$ given by~(\ref{ddrate4}). The scale $\mu_b$ is the matching scale of the order of the $b$-quark mass $m_b$.

%
%
\section{Coefficients $C^{(1)}_{i\,k;j}(m_b/\mu,\,r)$\label{sec:12}}

Here we give the explicit expressions for the coefficients $C^{(1)}_{i\,k;j}(m_b/\mu,\,r)$ (see~(\ref{wils:4})). 
To simplify the output we introduce functions $F_1$, $F_2$, and $F_3$:
\beqn
&&F_1(r,\,\mu/m_b)=-\frac{2}{3} \text{Log}\left[\frac{\mu ^2}{m_b^2 \left(1-4 r^2\right)^2}\right]^2-\frac{10}{3} \text{Log}\left[\frac{\mu^2}{m_b^2}\right]-\frac{\pi^2}{9}+\frac{8}{3} {\rm Li}_2\left[1-\frac{1}{1-4 r^2}\right]+\frac{4}{3} \text{Log}\left[1-4 r^2\right]^2,\nn
&&F_2(r)=\frac{\left(1-4 r^2\right) }{\left(1-2 r^2\right)}\left(\text{Log}\left[1-4 r^2\right]-i \pi \right),\nn
&&F_3(r,\,\mu/m_b)=-\frac{2}{3} \text{Log}\left[\frac{\mu ^2}{m_b^2}\right]^2+
\left(\frac{8}{3} \text{Log}\left[1-4 r^2\right] -\frac{6 \text{Log}[2 r] }{1-4 r^2}-\frac{13}{3}\right)\text{Log}\left[\frac{\mu ^2}{m_b^2}\right]-3 i \pi  \text{Log}\left[\frac{\left(2-4 r^2\right) \mu ^2}{\left(1-4 r^2\right)^2m_b^2 }\right]\nn
&&\qquad
+\frac{3 \left(1+4 r^2\right) }{-1+4 r^2}
\left({\rm Li}_2\left[\frac{1}{2-4 r^2}\right]-{\rm Li}_2\left[\frac{2 r^2}{1-2 r^2}\right]\right)-3 {\rm Li}_2\left[\frac{2 r^2}{-1+4 r^2}\right]-\frac{1}{3} {\rm Li}_2\left[\frac{4 r^2}{-1+4 r^2}\right]\nn
&&\qquad+\frac{17 \pi ^2}{9}-\frac{13}{3} \text{Log}\left[1-4 r^2\right]^2+3 \text{Log}\left[1-4 r^2\right] \text{Log}\left[1-2 r^2\right]
-\frac{12 \text{Log}[2 r] \text{Log}\left[1-4 r^2\right]}{-1+4 r^2}\nn
&&\quad+\frac{12 \text{Log}[r] \text{Log}\left[1-2 r^2\right]}{-1+4 r^2}
+\frac{3 \left(3+4 r^2\right) \text{Log}[2] \text{Log}\left[1-2 r^2\right]}{-1+4 r^2}+\frac{6 \text{Log}[r]^2}{1-4 r^2}+\frac{3 \left(3+4 r^2\right)}{2 \left(-1+4 r^2\right)}\text{Log}[2]^2.\nn
\eeqn

The coefficients $C^{(1)}_{i\,k;j}$ are enumerated according to the effective theory Dirac structures $j=1,2,3$.
\beqn
&&C^{(1)}_{00;1}=\frac{1}{2}\left(F_1(r,\,\mu/m_b)-\frac{56}{3}+\frac{8}{3} \text{Log}\left[1-4 r^2\right]\right),\nn
&&C^{(1)}_{08;1}=
\frac{1}{2}\left(-6 \text{Log}\left[\frac{\mu ^2}{m_b^2}\right]+\frac{3-8r^2}{1-2r^2}F_2(r)-23+\frac{1+2r^2}{1-2 r^2}\right),\nn
&&C^{(1)}_{80;1}=
\frac{2}{9}C^{(1)}_{08;1},\nn
&&C^{(1)}_{88;1}=
\frac{1}{2}\left(F_3(r,\,\mu/m_b)-\frac{i \pi  \left(21-104 r^2+152 r^4\right)}{6 \left(1-2 r^2\right)^2}\right.\nn
&&\qquad\qquad\left.+\frac{19-96 r^2+144 r^4}{6 \left(-1+2 r^2\right)^2}\text{Log}\left[1-4 r^2\right]
-3\text{Log}[4r]-\frac{32}{3}+\frac{7}{3 \left(1-2 r^2\right)}\right).\nn
&&C^{(1)}_{00;2}=-\frac{1}{4r}\left(F_1(r,\,\mu/m_b) -\frac{40}{3}+\frac{2 \text{Log}\left[1-4 r^2\right]}{3 r^2}\right),\nn
&&C^{(1)}_{08;2}=-\frac{1}{4r}\left(-6 \text{Log}\left[\frac{\mu ^2}{m_b^2}\right]+\frac{\left(3-4 r^2\right)}{\left(1-2 r^2\right)}F_2(r)-\frac{2 \left(11-20
r^2\right)}{1-2 r^2}\right),\nn
&&C^{(1)}_{80;2}=\frac{2}{9}C^{(1)}_{08;2},\nn
&&C^{(1)}_{88;2}=
-\frac{1}{4r}\left(F_3(r,\,\mu/m_b)-\frac{i \pi  \left(21-76 r^2+40 r^4\right)}{6 \left(1-2 r^2\right)^2}+\frac{27-40 r^2}{-3+6 r^2}-3 \text{Log}[4r]\right.\nn
&&\qquad\qquad\left.+\frac{-1+46 r^2-156 r^4+80 r^6}{12 r^2 \left(1-2 r^2\right)^2}\text{Log}\left[1-4 r^2\right]\right).\nn
&&C^{(1)}_{00;3}=-\frac{1}{2}\left(F_1(r,\,\mu/m_b)-\frac{56}{3}+\left(4-\frac{1}{3 r^2}\right) \text{Log}\left[1-4 r^2\right]\right),\nn
&&C^{(1)}_{08;3}=
-\frac{1}{2}\left(-6\text{Log}\left[\frac{\mu ^2}{m_b^2}\right]+F_2(r)-20-\frac{4 \text{Log}[2]}{1-2 r^2}-\frac{8 r^2 \text{Log}\left[r^2\right]}{1-2
r^2}\right),\nn
&&C^{(1)}_{80;3}=\frac{2}{9}C^{(1)}_{08;3},\nn
&&C^{(1)}_{88;3}=-\frac{1}{2}
\left(F_3(r,\,\mu/m_b)-6-\frac{i \pi  \left(7+8 r^2\right)}{6 \left(1-2 r^2\right)}-\frac{14 \text{Log}[2]}{3-6 r^2}+\frac{\left(-9+74 r^2\right) }{-3+6r^2}\text{Log}[r]\right.\nn
&&\qquad\qquad\left.-\frac{1+14 r^2+56 r^4}{24 r^2 \left(-1+2 r^2\right)} \text{Log}\left[1-4 r^2\right]\right).
\eeqn

%
%
\section{Transformation of operator basis\label{sec:9}}

In this section we present without derivation the relations used to calculate the Wilson coefficients $C_0^{NLO}(m_b)$ and $C_8^{NLO}(m_b)$ given in Eq.~(\ref{eft:7}). The Wilson coefficients for the set of operators 
\beq
\label{misyak}
  P_1 = [\bar{s} \gamma_\mu P_L T^a c] [\bar{c} \gamma^{\mu} P_L T^a b], \qquad
  P_2 = [\bar{s} \gamma_\mu P_L c] [\bar{c} \gamma^{\mu} P_L b].
\eeq
have been calculated in~\cite{Gorbahn:2004my} at the NNLO in $\alpha_s$. We need however the Wilson coefficients  in the different operator basis
\beq
\label{our}
  \Op_1 \equiv [\bar{s} \gamma_\mu P_L T^a b] [\bar{c} \gamma^{\mu} P_L T^a c], \qquad
  \Op_2 \equiv [\bar{s} \gamma_\mu P_L b] [\bar{c} \gamma^{\mu} P_L c].
\eeq
At the LO the two bases are related by a linear transformation which follows when one applies the Fierz transformations in the color and spinor spaces:
\beq
  \left(\begin{array}{c} \Op_1 \\ \Op_2 \end{array} \right) 
   = \left(\begin{array}{cc} -\frac{1}{3} & \frac{4}{9} \\ 2 & \frac{1}{3} \end{array} \right)
    \left(\begin{array}{c} P_1 \\ P_2 \end{array} \right).
   \label{trafo:d4}
\eeq

At the NLO the one-loop corrections have to be taken into account and that brings complications due to the evanescent operators. The dimensional regularization used when calculating Feynman integrals does not have an unambiguous definition for matrix $\gamma_5$, which is essentially a four-dimensional object. The systematic way of coping with this difficulty (see, e.g.~\cite{pa:evanescent}) consists in expanding the operator basis to include operators vanishing in the limit $D=4$ (evanescent operators). In our case the ambiguity is brought by the Fiertz transformation~(\ref{trafo:d4}) which is not defined at $D\neq4$. Working out the relation between the bases~(\ref{misyak}) and~(\ref{our}) at the NLO requires introduction of two evanescent operators 
\beq
 \left(\begin{array}{c} F_1 \\ F_2 \end{array} \right) =
  \left(\begin{array}{c} \Op_1 \\ \Op_2 \end{array} \right) 
   - \left(\begin{array}{cc} -\frac{1}{3} & \frac{4}{9} \\ 2 & \frac{1}{3} \end{array} \right)
    \left(\begin{array}{c} P_1 \\ P_2 \end{array} \right)
   \label{trafo:d41}
\eeq
and calculating one-loop diagrams analogous to those shown in Fig.~\ref{fig:FT} with the operators~(\ref{trafo:d41}) inserted. It suffices to extract only the UV-divergent part of the additional diagrams.

The Wilson coefficients of operators~(\ref{misyak}) and~(\ref{our}) are related by
\beq
  \label{WC:trafo:mu}
  \vec{C}'(\mu) = 
    \left( \begin{array}{cc} -\frac{1}{3} & 2 \\ \frac{4}{9} & \frac{1}{3} \end{array} \right)
    \left[ \left( \begin{array}{c} C_1^{(0)}(\mu) \\ C_2^{(0)}(\mu) \end{array} \right)
     + \frac{\alpha_s(\mu)}{4 \pi}  \left( \begin{array}{c} C_1^{(1)}(\mu) \\ C_2^{(1)}(\mu) \end{array} \right)
    \right]
  + \frac{\alpha_s(\mu)}{4 \pi}
    \left( \begin{array}{cc} 0 & 6 \\ -\frac{4}{3} & 0 \end{array} \right)
    \left( \begin{array}{c} C_1^{(0)}(\mu) \\ C_2^{(0)}(\mu) \end{array} \right).
\eeq
Here $\vec{C}'(\mu)$ are the Wilson coefficients for the basis~(\ref{our}) and $\vec{C}(\mu)$ for the basis~(\ref{misyak}). The first matrix in Eq.~(\ref{WC:trafo:mu}) is the transposed matrix in Eq.~(\ref{trafo:d4}) corresponding to the tree-level transformation of the operator basis. The second matrix in~(\ref{WC:trafo:mu}) is the contribution of operators~(\ref{trafo:d41}).

The explicit expressions for the Wilson coefficients $\vec{C}(\mu)$ can be found in Eqs.~(39)-(44) in~\cite{Gorbahn:2004my}. Substituting them into Eq.~(\ref{WC:trafo:mu}) gives the required expressions for  $C_0^{NLO}(m_b)$ and $C_8^{NLO}(m_b)$.

%
%
\section{Dirac structures of the effective theory\label{sec:10}}

In this section we present the reduction formulae necessary to project
the matrix elements of the full theory into the matrix elements of the effective theory. The generic Dirac structure on the full QCD side has the form
\begin{equation}
  \label{FTME}
  \left[ \bar{s}\, \Gamma_1\, b \right] \left[ \bar{c}\, \Gamma_2\, c \right],
\end{equation}
where $\Gamma_{1,2}$ is one of the following products of Dirac matrices
\begin{equation}
  \label{set}
  \Gamma = \left\{ 1;\,\g_5;\, \g^\mu ;\,\g^\mu\g_5;\,\g^\mu\g^\nu;\, 
           \g^\mu\g^\nu\g_5;\, \g^\mu\g^\nu\g^\eta;\, \g^\mu\g^\nu\g^\eta\g_5\right\}.
\end{equation}
Here the last four entries come from one-loop corrections to the
tree-level full theory operator. On the effective theory side the generic matrix element is
\begin{equation}
  \label{ETME}
  [\bar{\xi}_{n,p}\Gamma^{ET}_{sb}\, h_{\beta_b}]
  [\bar{\xi}_{\beta_{\cc}\,\pp}\Gamma^{ET}_{\cc}\,\eta^C_{\beta_{\cc}\,-\pp}],
\end{equation}
where $\Gamma^{ET}_{sb}$ and $\Gamma^{ET}_{\cc}$ are the set of
operators which form the complete set on the corresponding subspace of
Dirac spinors.

The Dirac structures $\Gamma^{ET}_{\cc}$ follow by sandwiching the full theory structures of Eq.~(\ref{set}) between the projectors $P_{\beta_{\cc}}$ and $P_{-\beta_{\cc}}$
\beq
  P_{\beta_{\cc}} \Gamma P_{-\beta_{\cc}}\rightarrow \Gamma^{ET}_{\cc}.
\eeq
Working out the Dirac structures $\Gamma^{ET}_{sb}$ requires more effort because now
the original Dirac structure from the list~(\ref{set}) is sandwiched between the projectors on different subspaces: $P_{\bar{n}}$ and $P_{\beta_b}$. The most straightforward way to work out the necessary relations is using Eq.~(17) from~\cite{pa:0211251} that reads
\beq
  \label{complete}
  \Gamma^{ET}_{sb}=\nb\,\tr\left[\n \Pnb \Gamma \Pb \right] 
  - \nb \g_5 \,\tr\left[\n \g_5 \Pnb \Gamma \Pb \right]
  +\g^\mu_\perp\,\tr \left[\g^\perp_\mu \Pnb \Gamma \Pb \right].
\eeq
Here $\g^\mu_\perp = \g^\mu - n^\mu \nb - \bar{n}^\mu \n$.

The projection formulae are considerably simplified in the restframe of the $b$-quark ($\bRF$) where $2\beta_b=n+\bar{n}$. The four-momentum of the $s$-quark in the $\bRF$ is specified by its energy: $p^\mu=E_s n^\mu$. To simplify the results further we use the relations
\beq
  \label{polarization}
\g^\mu_\perp - i\eps^{\mu\nu}_\perp\g_{\nu\,\perp} = -2\eps^\mu_+\dslash{\eps}_-,
  \qquad  
  \g^\mu_\perp + i\eps^{\mu\nu}_\perp\g_{\nu\,\perp} = -2\eps^\mu_-\dslash{\eps}_+,
 \eeq
where $\eps_\pm$ are the polarization vectors of $s$-quark with $\pm$ corresponding
to the spin direction along/opposite to the quark momentum. Here $\eps^{\mu\nu}_\perp=\eps^{\mu\nu\eta\rho}\beta_{b\,\eta} n_\rho$. These relations can be verified explicitly in the $\bRF$ where $n^\mu=(1,0,0,1)$, $\bar{n}^\mu=(1,0,0,-1)$, $\beta^\mu_b=(1,0,0,0)$, and $\eps^\mu_\pm=2^{-1/2}(0,1,\pm i,0)$. Equations below summarize the results.

On the $\cc$-side the following relations hold:
\beqn
   \label{ccbarR}
   \otimes \g^\mu P_L&&\rightarrow \otimes 
   \frac{1}{2}\g^\mu_\perp - \otimes \frac{1}{2}\beta^\mu_{\cc}\g_5, \nn
    \otimes \g^\mu P_R&&\rightarrow \otimes 
   \frac{1}{2}\g^\mu_\perp + \otimes \frac{1}{2}\beta^\mu_{\cc}\g_5, \nn
   \otimes \g^{[\mu}\g^{\nu]}P_L&&\rightarrow \otimes
   \frac{1}{2}(\beta_{\cc}^\mu \g^\nu_\perp - \beta_{\cc}^\nu \g^\mu_\perp) 
   - \otimes \frac{i}{2}\eps^{\mu\nu\rho\eta}\beta_{\cc\,\rho}\g_{\eta _\perp},\nn
    \otimes \g^{[\mu}\g^{\nu]}P_R&&\rightarrow \otimes
   \frac{1}{2}(\beta_{\cc}^\mu \g^\nu_\perp - \beta_{\cc}^\nu \g^\mu_\perp) 
   + \otimes \frac{i}{2}\eps^{\mu\nu\rho\eta}\beta_{\cc\,\rho}\g_{\eta _\perp}.
\eeqn
Here $\g^\mu_\perp=\g^\mu-  \beta^\mu\dslash{\beta}$ and $\g^{[\mu}\g^{\nu]}= \frac{1}{2}[\g^\mu,\g^\nu]$. On the $\bar{s}b$-side we use relations
\beqn
     \label{sbR}
     \g^\mu P_L\otimes&&\rightarrow n^\mu P_R\otimes - \eps^\mu_+\dslash{\eps}_-\otimes,\nn
     \g^\mu P_R\otimes&&\rightarrow n^\mu P_L\otimes - \eps^\mu_-\dslash{\eps}_+\otimes,\nn
     \g^{[\mu}\g^{\nu]} P_L\otimes &&\rightarrow (n^\mu\beta_b^\nu - n^\nu\beta_b^\mu) P_L\otimes
     +i\eps^{\mu\nu}_\perp P_L\otimes
     +(n^\mu \eps^\nu_- - n^\nu \eps^\mu_-)\dslash{\eps}_+\otimes,\nn
      \g^{[\mu}\g^{\nu]} P_R\otimes&&\rightarrow (n^\mu\beta_b^\nu - n^\nu\beta_b^\mu) P_R\otimes
     -i\eps^{\mu\nu}_\perp P_R\otimes
     +(n^\mu \eps^\nu_+ - n^\nu \eps^\mu_+)\dslash{\eps}_-\otimes.
\eeqn

Using Eqs.~(\ref{ccbarR}) and~(\ref{sbR}) it is straightforward to verify~(\ref{ETstruct1}). Finally, there is one more relation holding at the leading order in $v$-expansion that allowed us to reduce the number of the effective theory Dirac structures when processing the output of the full QCD calculation: 
\beq
\label{czero}
\otimes \dslash{\beta}_{b\,\perp}=\frac{E_s}{m_b}\otimes \dslash{n}_{\perp}.
\eeq 
It follows from conservation of momentum, $m_b\beta_b=2 m_c\beta_{\cc}+E_s n$, and observation that on the $\cc$-side only the components perpendicular to $\beta_{\cc}$ contribute, which is obvious from~(\ref{ccbarR}).

%

%

\end{document}